\documentclass[12pt]{article}
\usepackage{amsmath}
\usepackage{amssymb} 
\usepackage{epsfig}
\usepackage{amsthm}
\usepackage{stmaryrd}
\usepackage{bm}
\usepackage{hyperref}
\usepackage{eufrak}
\usepackage{upgreek}

\numberwithin{equation}{section}
\numberwithin{figure}{section}

\def\eq#1{(\ref{eq:#1})}
\def\lineup{\!\!\!\!\!\!\!\!&&}

\newcommand{\Tr}{\mathop{\rm Tr}\nolimits}
\newcommand{\?}{\stackrel{?}{=}}
\def\d{\partial}
\def\eps{\epsilon}

\def\fraction#1#2{ { \textstyle \frac{#1}{#2} }}
\def\half{\fraction{1}{2}}

\def\s{\sigma}
\def\sb{\overline{\sigma}}
\def\BCFT{\mathrm{BCFT}}
\def\tv{\mathrm{tv}}
\def\Sch{\mathrm{Sch}}
\def\ND{\mathrm{ND}}
\def\S{\Sigma}
\def\Sb{\overline{\Sigma}}

\begin{document}

\begin{titlepage}
\hfill LMU-ASC 39/14
 
\begin{center}

\vskip .5cm {\large \bf{String Field Theory Solution for 
Any Open String Background}}

\vskip 1.5cm

{\large Theodore Erler$^{a,}$\footnote{tchovi@gmail.com}, 
Carlo Maccaferri$^{b,c,}$\footnote{maccafer@gmail.com}}

\vskip 1.0cm

$^{a}${\it Arnold Sommerfeld Center for Theoretical Physics,}\\
{\it Theresienstrasse 37, D-80333, Munich, Germany}
\vskip .5cm

$^b${\it Dipartimento di Fisica, Universit\'a di Torino and INFN, Sezione di 
Torino,}\\
{\it Via Pietro Giuria 1, I-10125 Torino, Italy}
\vskip .5cm

$^c${\it Institute of Physics of the ASCR, v.v.i.}\\
{\it Na Slovance 2, 182 21 Prague 8, Czech Republic}

\vskip 1.5cm

{\bf Abstract}
\end{center}

We present an exact solution of open bosonic string field theory which can be
used to describe any time-independent open string background. The solution 
generalizes an earlier construction of Kiermaier, Okawa, and Soler, and 
assumes the existence of boundary condition changing operators with 
nonsingular OPEs and vanishing conformal dimension. Our main observation is 
that boundary condition changing operators of this kind can describe nearly 
any open string background provided the background shift is accompanied by 
a timelike Wilson line of sufficient strength. As an 
application we analyze the tachyon lump describing the formation of a 
D$(p-1)$-brane in the string field theory of a D$p$-brane, for generic 
compactification radius. This not only provides a proof of Sen's second 
conjecture, but also gives explicit examples of higher energy solutions, 
confirming analytically that string field theory can ``reverse'' the 
direction of the worldsheet RG flow. We also find multiple D-brane solutions, 
demonstrating that string field theory can add Chan-Paton factors and 
change the rank of the gauge group. Finally, we show how the solution 
provides a remarkably simple and nonperturbative proof of the background 
independence of open bosonic string field theory.

\noindent

\noindent
\medskip

\end{titlepage}

\tableofcontents

\section{Introduction}

Following Schnabl's analytic solution for tachyon condensation 
\cite{Schnabl}, analytic techniques in open string field theory have 
provided a remarkably clear and beautiful description of the endpoint of 
tachyon condensation on unstable D-branes \cite{cohomology,simple,BerkVac}. 
However, efforts to extend these techniques beyond the universal sector 
have been less fruitful.  Several solutions describing marginal 
deformations have been found, especially as a perturbative expansion in 
the deformation parameter 
\cite{KROZ,Sch_marg,FK,KO,sKROZ,sKROZ2,sKROZ3,sFK,sKO}.\footnote{A more 
nonperturbative approach to marginal deformations was provided by the KOS
solution \cite{KOS}, which will play a central role in our discussion, and 
the old identity-based solution for marginal deformations introduced by 
Takahashi and Tanimoto \cite{TT}, for which there have been interesting 
recent developments \cite{TTbos,TTbos2,TTcub,TTsup}. A solution which 
aims to unify these approaches was recently proposed in \cite{Macc}.} 
But the main question about marginal deformations is whether string field 
theory can describe the full moduli space of vacua connected to a given 
D-brane system \cite{marg_trunc,Sen_marg,Kaczmarek,marg_trunc2}, and this \
question seems out of reach in a perturbative approach. 
On a different line of thought, there have also been interesting proposals
to describe the formation of lower dimensional D-branes by following a 
given boundary world-sheet RG flow \cite{Ellwood,BMT}. A success was the 
computation of the energy and closed string tadpole \cite{BMT,Bonora,lumps}, 
but further work has encountered subtle problems with the equation of motion 
\cite{Bonora,lumps,Integra,Bonora2}, and has been limited by the very few 
known soluble worldsheet RG-flows. 

In light of these difficulties, one particularly attractive proposal was 
advanced by Kiermaier, Okawa, and Soler (KOS) \cite{KOS}. By 
making a gauge transformation of the solution \cite{KROZ,Sch_marg} for 
nonsingular marginal deformations, they managed to construct a solution 
which could be expressed directly in terms of boundary condition changing 
(bcc) operators $\s,\sb$ relating the perturbative vacuum to the boundary 
conformal field theory (BCFT) of the D-brane system one wishes to describe. 
Since the existence of bcc operators relating BCFTs is a generic fact, this 
suggests a kind of all-purpose string field theory solution which could 
be used to describe any open string background. For the
KOS solution to work, however, the bcc operators must
satisfy a rather unusual property:
\begin{equation}\lim_{s\to 0}\sb(s)\s(0)=1,\ \ \ \ \ (s>0).
\label{eq:reg}\end{equation}
While this is satisfied for backgrounds related by nonsingular marginal 
deformations, usually bcc operators have nonvanishing conformal weight, 
and their OPEs are singular. Efforts to generalize the KOS solution to 
avoid \eq{reg} have so far been unsuccessful.

In this paper we observe that bcc operators satisfying \eq{reg} 
can, in fact, describe any change of boundary condition provided the 
time component of the $X^\mu$ BCFT is unaltered. The idea is as follows.
Suppose $\s_{*},\sb_{*}$ are bcc operators of weight $h$ satisfying
\begin{equation}\sb_{*}(s)\s_{*}(0)\sim 
\frac{1}{s^{2h}}+\mathrm{less\ singular},\ \ \ \ \ (s>0),
\end{equation}
and which act as the identity operator in the time direction. We will 
construct an analytic solution using a modified pair of bcc 
operators\footnote{We use $\alpha'=1$ units.}
\begin{equation}\s(s)=\s_* e^{i \sqrt{h}X^0}(s),\ \ \ \ \ 
\sb(s)=\sb_* e^{-i \sqrt{h}X^0}(s).\label{eq:s_sb}\end{equation}
The plane-wave factors $e^{\pm i\sqrt{h}X^0}$ cancel the conformal weight of 
$\s_*$ and $\sb_*$, and because
\begin{equation}e^{-i\sqrt{h}X^0}(s)e^{i\sqrt{h}X^0}(0)\sim s^{2h},\ \ \ \ \ 
(s>0),
\end{equation}
the modified bcc operators satisfy equation \eq{reg}. The resulting solution
will have nontrivial primaries excited in the $X^0$ BCFT, which {\it a priori}
could effect the physical interpretation of the solution. In fact, 
the $e^{\pm i\sqrt{h}X^0}$ factors are bcc operators which turn on a 
Wilson line in the time direction. But since the only physical effect of 
a Wilson line is through winding modes, and the time direction is 
noncompact, the timelike Wilson line is physically invisible. 
In field theory, a constant timelike Wilson line is pure gauge:
\begin{equation}A_\mu = \lambda \delta_\mu^{\ 0} = e^{i \lambda x^0}
i\d_\mu(e^{-i \lambda x^0}),
\end{equation}
which suggests that the timelike primaries excited by $\s,\sb$ could 
likewise be removed by a gauge transformation in string field theory, though
doing this in practice may be nontrivial. 

The implications of this simple idea are profound. It means that string field
theory can provide a closed form description of a far greater range of 
backgrounds than have been identified in level truncation or analytically, 
and in fact the bulk of D-brane setups one might care to consider in string 
theory. After modest generalization of the considerations of KOS, the 
solution is extraordinarily simple. Finding the energy, the closed string 
tadpole, and the cohomology are easily reduced to worldsheet computations. 
Remarkably, the solution even satisfies (a generalization of \cite{simple}) 
the Schnabl gauge condition.

\section{Algebra}
\label{sec:alg}

We begin by quickly reviewing the algebraic ingredients we need to 
formulate the solution. This is (mostly) standard material; See also the 
original paper by KOS \cite{KOS} and Noumi and Okawa 
\cite{sKOS},\footnote{We follow the 
conventions of \cite{simple}, in particular we use the ``left-handed'' 
star product. KOS use the ``right-handed'' star product, and the opposite 
sign for the fields $K$ and $B$.} and for further explanations of the 
algebraic formalism we use, see \cite{simple,Okawa,SSF1,OkawaRev}. 
 
We start with string field theory formulated around some reference D-brane 
system, described by a boundary conformal field theory $\BCFT_0$. Then
we construct a classical solution describing some other D-brane 
system, described by a boundary conformal field theory $\BCFT_*$. We 
assume that $\BCFT_0$ and $\BCFT_*$ are factorized in the form
\begin{equation}
\BCFT_{c=25}\otimes\BCFT_{X^0}\otimes\BCFT_{bc}\label{eq:bkgd}
\end{equation}
$\BCFT_0$ and $\BCFT_*$ share a common $bc$ ghost factor and a noncompact, 
timelike free boson $X^0$ subject to Neumann boundary conditions. The $c=25$ 
components of the two BCFTs can be different and essentially arbitrary provided
they share the same bulk CFT. In this way, the shift between the backgrounds
$\BCFT_0$ and $\BCFT_*$ can be represented by boundary condition changing 
operators, as explained in the introduction. For backgrounds not of the form 
\eq{bkgd} we do not have a general construction, though in some cases such 
backgrounds can be realized.\footnote{The exponential timelike 
\cite{KROZ,Sch_marg,KOS,Senroll,rolling} and lightlike 
\cite{HS,Ghoshal} rolling tachyon solutions are examples of nonsingular 
marginal deformations, and therefore are described by boundary condition 
changing operators of the kind needed to construct the solution. The 
$\cosh(X^0)$ deformation \cite{Senroll} could be realized by turning on an 
imaginary Wilson line in a noncompact spacelike direction, if available. These 
backgrounds are not described by BCFTs of the form \eq{bkgd}.}

\begin{figure}
\begin{center}
\resizebox{6in}{2.1in}{\includegraphics{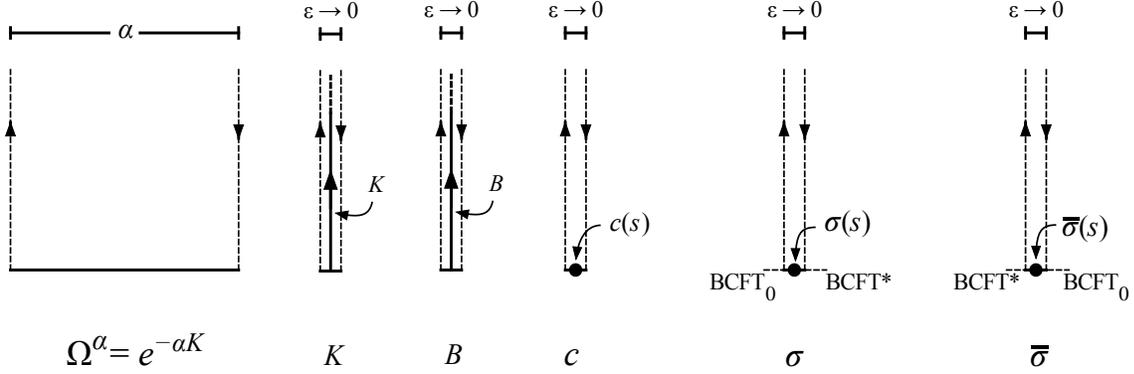}}
\end{center}
\caption{\label{fig:KOSing1} The wedge state $\Omega^{\alpha}$ and the 
fields $K,B,c,\s,\sb$ represented as semi-infinite ``strips'' with operator 
insertions in correlation functions on the cylinder. Note that star 
multiplication of two string fields glues the right half of the first strip
to the left half of the second strip.}
\end{figure}

The solution is formulated within the subalgebra of wedge states with 
operator insertions \cite{Schnabl,Schnabl_wedge}. A wedge state 
\cite{RZ_wedge} is any positive star algebra power 
of the $SL(2,\mathbb{R})$ vacuum $\Omega\equiv|0\rangle$:
\begin{equation}\Omega^\alpha,\ \ \ \ \alpha\geq 0.
\end{equation}
Here (and in the rest of the paper) we omit the $*$ symbol when multiplying 
string fields. In the limit $\eps\to 0$, the wedge state $\Omega^\eps$ 
approaches the formal identity of the star algebra, called the identity 
string field. We write the identity string field simply as $1$. The conformal 
field theory definition of a wedge state is easiest to visualize in the 
sliver coordinate frame \cite{Schnabl,Okawa,bef,RZO}, where 
$\Omega^\alpha$ is represented as a semi-infinite, vertical ``strip'' of 
worldsheet of width $\alpha$, as shown in figure \ref{fig:KOSing1}. The 
``strip'' can be glued to itself or to other ``strips'' along the vertical 
edges, resulting in worldsheet correlation functions on the cylinder (which 
can be mapped to the upper half plane). To describe the solution, the 
``strips'' should also contain particular operator insertions, specifically, 
boundary insertions of the $c$-ghost, 
\begin{equation}c(s),\end{equation}
boundary condition changing operators,
\begin{equation}\s(s),\ \ \ \sb(s),\end{equation}
and vertical line integral insertions of the energy-momentum tensor and
$b$-ghost,
\begin{eqnarray}
K\lineup = \int_{-i\infty}^{i\infty}\frac{dz}{2\pi i} T(z),\\
B\lineup = \int_{-i\infty}^{i\infty}\frac{dz}{2\pi i} b(z).
\end{eqnarray} 
The operator $\s$ changes the open string boundary condition from 
$\BCFT_0$ to $\BCFT_*$, and $\sb$ changes the boundary condition in reverse, 
from $\BCFT_*$ back to $\BCFT_0$. We assume that $\s$ 
and $\sb$ are weight zero primaries constructed by tensoring a primary 
bcc in the $c=25$ component of the $\BCFT$ with a timelike Wilson line. 
However, much of our discussion can be generalized to non-primary bccs. 

This class of states can be conveniently expressed by taking star products 
of five ``atomic'' string fields:
\begin{equation}K,\ B,\ c,\ \s,\ \sb.\label{eq:fields}\end{equation}
Each string field can be defined as an infinitesimally thin ``strip'' 
carrying the respective operator insertion (denoted by the same symbol), 
as shown in figure \ref{fig:KOSing1}.\footnote{Note that $\s,\sb$ are 
fields of a stretched string between $\BCFT_0$ and $\BCFT_*$. They 
are not string fields in $\BCFT_0$.} The field $K$ generates the algebra 
of wedge states, in that any positive power of the $SL(2,\mathbb{R})$ 
vacuum can be written
\begin{equation}\Omega^\alpha = e^{-\alpha K}.\end{equation}
Of particular importance are the string fields \cite{simple}
\begin{eqnarray}
\frac{1}{1+K} \lineup = \int_0^\infty d\alpha\, e^{-\alpha}\Omega^\alpha,\\
\frac{1}{\sqrt{1+K}}\lineup = \frac{1}{\sqrt{\pi}}\int_0^\infty d\alpha
\frac{e^{-\alpha}}{\sqrt{\alpha}}\Omega^\alpha,
\end{eqnarray}
which are defined via the Schwinger parameterization in terms of a 
continuous superposition of wedge states. Multiplying a wedge state with 
$K,B,c,\s,\sb$ on the left (right) effectively inserts the corresponding 
operator on the left (right) edge of the ``strip'' defining the wedge state. 
In this way, we can insert all of the operators we need by taking star 
products of these five basic string fields.

The fields satisfy a number of algebraic relations. First, we have 
BRST variations:
\begin{equation}
QK=0;\ \ \ QB=K;\ \ \ Qc = c\d c;\ \ \ Q\s = c\d \s;\ \ \ Q\sb = c\d \sb,
\label{eq:BRST}\end{equation}
where $\d \equiv [K,\cdot]$. Note that the BRST variation of $\s$ and $\sb$ is 
exactly like that of a dimension zero matter primary. We also have 
algebraic relations:
\begin{eqnarray}
\lineup B^2=c^2=0;\ \ \ [K,B]=0;\ \ \ \ \ \ \, Bc+cB=1;\nonumber\\
\lineup [\s,c]=0;\ \ \ \ \ \ \ \,[\s,\d c]=0;\ \ \ \ \ \ \ [\s,B]=0;\nonumber\\
\lineup [\sb,c]=0;\ \ \ \ \ \ \ \,[\sb,\d c]=0;\ \ \ \ \ \ \ [\sb,B]=0.
\label{eq:alg}\end{eqnarray}
The first three relations are well-known \cite{Okawa}, and the last six
follow trivially from the fact that $\s$ and $\sb$ represent matter 
operators. Finally, we have two important relations:
\begin{eqnarray}
\sb\s \lineup = 1;\label{eq:sbs}\\
\s\sb \lineup = \mathrm{finite}.\label{eq:ssb}
\end{eqnarray}
The first equation follows from the OPE \eq{reg} discussed in the 
introduction. The second equation is somewhat surprising, since if 
$\s\sb\neq 1$ we have an ``associativity anomaly:''
\begin{equation}(\s\sb)\s \neq \s(\sb\s).\label{eq:ass}\end{equation}
This reflects an ambiguity in the definition of correlators when 3 or 
more bccs collide. (See appendix \ref{app:twist}.) This leads to a few 
subtleties which are important to be aware of. But in practice the 
product $\s\sb$ never appears in all essential computations with the 
solution, and there is no need to assign it a definite value. So 
associativity anomalies do not appear. 

Still, it is important to understand why $\s\sb\neq 1$ in general.
Consider the 2-point function of bcc operators on the unit disk:
\begin{equation}\langle \sb(1)\s(e^{i\theta})\rangle.\end{equation}
For angles in the range $[0,\theta]$ the correlator has $\BCFT_0$ boundary
conditions, and outside this range it has $\BCFT_*$ boundary conditions. 
Since $\s$ and $\sb$ are dimension zero primaries, this 2-point function is 
independent of the angular separation. Therefore, in the limit $\theta\to 0^+$ 
we can use the OPE \eq{reg} to find
\begin{equation}\langle \sb(1)\s(e^{i\theta})\rangle=g_*,\end{equation}
where $g_*$ is the disk partition function in $\BCFT_*$ (the $g$-function). 
Now consider $\theta\to 2\pi^-$. In this limit the correlator should be 
proportional to $g_0$---the disk partition function in $\BCFT_0$---times 
the coefficient of the identity operator in the $\s$-$\sb$ OPE. But since the 
correlator must be equal to $g_*$, we find 
\begin{equation}\lim_{s\to 0} \s(s)\sb(0) = \frac{g_*}{g_0},\ \ \ \ \ (s>0).
\end{equation}
The disk partition functions will be different if the D-brane configurations
have different energies. So in general $\s\sb\neq 1$.

Let us explain another puzzle, which in the past seemed to give a compelling 
argument that the KOS solution could only describe marginal 
deformations. Since $\s$ and $\sb$ are 
weight zero primaries with regular OPE, it is natural to define the operator
\begin{equation}V=\s\d\sb.\label{eq:V}\end{equation}
This should be a weight 1 primary, and it naively defines a 1-parameter 
family of conformal boundary conditions connecting $\BCFT_0$ and 
$\BCFT_*$. To see this, consider a wedge state deformed by a $V$-boundary 
interaction \cite{KOS}
\begin{equation}e^{-(K+\lambda V)}.\label{eq:def}\end{equation}
At $\lambda=0$, this describes the boundary condition of $\BCFT_0$. Meanwhile,
at $\lambda=1$ (assuming $\s\sb=1$),
\begin{eqnarray}
e^{-K-\s\d\sb} = e^{-K-\s K\sb+\s\sb K} = e^{-\s K\sb} = \s \Omega \sb.
\label{eq:farg}
\end{eqnarray}
So we find the boundary condition of $\BCFT_*$. Thus it seems that $\BCFT_*$
must represent a marginal deformation of $\BCFT_0$. The problem is that this 
argument makes assumptions about the nature of 
short distance collisions of the $\s$s and $V$ which are not valid in 
general. We will explain how this happens (for lump solutions) in 
section \ref{sec:lumps}.

\section{Solution}

The solution is most easily described by starting with string field theory 
formulated around the tachyon vacuum. Specifically, we begin with the 
``simple'' solution for the tachyon vacuum, introduced in 
\cite{simple}:
\begin{equation}\Psi_\tv = \frac{1}{\sqrt{1+K}}c(1+K)Bc\frac{1}{\sqrt{1+K}}.
\label{eq:simple}\end{equation}
The equations of motion expanded around the tachyon vacuum are
\begin{equation}Q_{\Psi_\tv}\Phi +\Phi^2 = 0,\label{eq:tvEOM}\end{equation}
where $Q_{\Psi_\tv} = Q+[\Psi_\tv,]$ is the shifted kinetic operator. To 
describe the perturbative vacuum $\BCFT_0$, we should take the solution
\begin{equation}\Phi=-\Psi_\tv.\end{equation}
Now suppose we want to describe some other D-brane system 
$\BCFT_*$. A natural guess would be to subtract the 
tachyon vacuum in $\BCFT_*$:
\begin{equation}\Phi \?-\s \Psi_\tv \sb,\label{eq:guess}\end{equation}
where $\s$ and $\sb$ are needed to translate the degrees of freedom of 
$\BCFT_*$ into $\BCFT_0$. Surprisingly, this almost works. It would be a 
solution if $Q_{\Psi_\tv}$ annihilated $\s$ and $\sb$. 

In this connection, it is worth noting that $\s$ and $\sb$ are killed by the 
kinetic operator of a different tachyon vacuum solution, namely, a singular 
tachyon vacuum consisting of a divergent insertion of the $c$ ghost 
\cite{cubic}:\footnote{Plugging into the equations of motion, this is a 
solution at order $\frac{1}{\alpha^2}$ since $c^2=0$. To get subleading 
orders to work requires a regularization of the solution.}
\begin{equation}\Psi_\mathrm{singular}= 
\frac{1}{\alpha} c,\ \ \ \ \alpha\to 0.
\label{eq:singtv}\end{equation}
This is closely related to the boundary string field theory description 
of the tachyon vacuum \cite{BSFT,Kutasov}, in that it naturally leads to 
a boundary deformation of the worldsheet action given by an infinite 
constant which sets all correlators to zero. The corresponding kinetic 
operator leaves $\s$ and $\sb$ invariant in a trivial way since 
$[\s,c]=[\sb,c]=0$. This suggests we should look for ``regularized'' 
analogues of $\s$ and $\sb$ which are left invariant by $Q_{\Psi_\tv}$:
\begin{equation}Q_{\Psi_\tv}\S=0;\ \ \ \ Q_{\Psi_\tv}\Sb = 0.\end{equation}
We can guess the needed expressions $\S$ and $\Sb$ as follows. Since 
$Q_{\Psi_\tv}$ has trivial cohomology, we should be able to write $\S$ and 
$\Sb$ as $Q_{\Psi_\tv}$ of some string field. If there is no change 
of boundary condition, we expect $\S=\Sb=1$, in which case this string 
field can be nothing but the homotopy operator for the tachyon vacuum 
\cite{cohomology,simple}, which satisfies
\begin{equation}1=Q_{\Psi_\tv}\left(\frac{B}{1+K}\right).\end{equation}
If the change of boundary condition is nontrivial, one might guess that 
the homotopy operator should be accompanied by an insertion of a bcc 
operator. Thus we are lead to the expressions\footnote{Closely related 
expressions appear in \cite{Macc,Ellwood} in the description of the 
cohomology for marginal deformations.}
\begin{eqnarray}
\S\lineup = Q_{\Psi_\tv}\left(\frac{1}{\sqrt{1+K}}B \s \frac{1}{\sqrt{1+K}}
\right);\label{eq:S}\\
\Sb \lineup 
=  Q_{\Psi_\tv}\left(\frac{1}{\sqrt{1+K}}B \sb \frac{1}{\sqrt{1+K}}\right).
\label{eq:Sb}
\end{eqnarray}
By construction, these fields are killed by $Q_{\Psi_\tv}$. To have a solution
to the equations of motion, $\S$ and $\Sb$ must satisfy the additional 
property 
\begin{equation}\Sb\S=1.\end{equation}
This can be demonstrated as follows:
\begin{eqnarray}
\Sb\S \lineup 
= Q_{\Psi_\tv}\left(\frac{1}{\sqrt{1+K}}\sb B\frac{1}{\sqrt{1+K}}\right)
Q_{\Psi_\tv}\left(\frac{1}{\sqrt{1+K}}B\s\frac{1}{\sqrt{1+K}}\right),
\nonumber\\
\lineup = Q_{\Psi_\tv}\left(\frac{1}{\sqrt{1+K}}\sb B\frac{1}{\sqrt{1+K}}
Q_{\Psi_\tv}\left(\frac{B}{1+K}\sqrt{1+K}\s\frac{1}{\sqrt{1+K}}\right)\right),
\nonumber\\
\lineup = Q_{\Psi_\tv}\left(\frac{1}{\sqrt{1+K}}\sb B\frac{1}{\sqrt{1+K}}
Q_{\Psi_\tv}\left(\frac{B}{1+K}\right)\sqrt{1+K}\s\frac{1}{\sqrt{1+K}}\right),
\nonumber\\ 
\lineup =  Q_{\Psi_\tv}\left(\frac{1}{\sqrt{1+K}}\sb B\frac{1}{\sqrt{1+K}}
\sqrt{1+K}\s\frac{1}{\sqrt{1+K}}\right),\nonumber\\
\lineup = Q_{\Psi_\tv}\left(\frac{B}{1+K}\right)=1.\label{eq:SSbproof}
\end{eqnarray}
Taking the product in the opposite order, we also 
have $\S\Sb = \frac{g_*}{g_0}$, so $\S$ and $\Sb$ multiply just like 
$\s$ and $\sb$.\footnote{Note that the derivation \eq{SSbproof} only 
requires $\sb\s = 1$ and $[B,\s]=[B,\sb]=0$. Therefore, all other relations 
satisfied by $\s$ and $\sb$ in equations \eq{BRST} and \eq{alg} are not 
needed to have a solution to the equations of motion. In particular 
$\s$ and $\sb$ do not necessarily have to be primaries, but in this case 
the solution will take a different form from \eq{sol}, and will 
not satisfy the gauge condition \eq{simp_gauge}. In this paper we have 
a specific realization of $\s$ and $\sb$ in mind, so we will assume all 
relations in \eq{BRST} and \eq{alg} without qualification.}

Therefore, after replacing $(\s,\sb)$ with $(\S,\Sb)$, our initial guess for 
the solution turns out to be correct: 
\begin{equation}\Phi = -\S \Psi_{\tv}\Sb,\label{eq:Phi}\end{equation}
Substituting previous expressions for $\Psi_\tv$, $\S$ and $\Sb$ we 
find more explicitly,
\begin{equation}
\Phi = -\frac{1}{\sqrt{1+K}}c(1+K)\s\frac{B}{1+K}
\sb(1+K)c\frac{1}{\sqrt{1+K}}.\label{eq:Phiexp}
\end{equation}
This is a solution to the equations of motion around the tachyon vacuum. 
Shifting back to the perturbative vacuum, the solution takes the form:
\begin{eqnarray}\Psi\lineup = \Psi_\tv-\S\Psi_\tv\Sb,\nonumber\\
\lineup = 
\frac{1}{\sqrt{1+K}}c\left[(1+K)-(1+K)\s \frac{1}{1+K}
\sb(1+K)\right]Bc\frac{1}{\sqrt{1+K}}.\label{eq:sol}
\end{eqnarray}
In the special case $\s\sb=1$, this expression is equivalent to the 
original solution proposed by KOS \cite{KOS}:
\begin{equation}\Psi_\mathrm{KOS} = -\frac{1}{\sqrt{1+K}}
c\d \sigma\frac{1}{1+K}\sb(1+K)Bc
\frac{1}{\sqrt{1+K}},\ \ \ \ (\s\sb=1).\label{eq:KOS}\end{equation}
To see this, use 
\begin{equation}\d\s = (1+K)\s - \s(1+K),\end{equation}
and plug in to the KOS solution to reproduce \eq{sol}. Note that when 
$\s\sb\neq 1$, the KOS solution does not satisfy the equations of motion, 
whereas \eq{sol} does.

Let us explain a few technical properties of the solution. It satisfies the 
string field reality condition,\footnote{The operation $^\ddag$ is defined 
as the composition of Hermitian and BPZ conjugation \cite{Tensor}. It is 
formally analogous to Hermitian conjugation of an operator. The fields $K,B$ 
and $c$ are self-conjugate, $K^\ddag=K$ etc, while $\s^\ddag = \sb$.}
\begin{equation}\Psi^\ddag = \Psi.\end{equation}
A nice property of the reality condition is that $\S$ and $\Sb$ are 
conjugate to each other: 
\begin{equation}\S^\ddag = \Sb;\ \ \ \ \Sb^\ddag = \S.\end{equation} 
and therefore are analogous to unitary operators. (The analogy is not 
complete because $\S\Sb\neq1$ in general). However, for some purposes it 
is natural to work with a non-real form of the solution \cite{simple}:
\begin{eqnarray}\Psi'\lineup =\sqrt{1+K}\Psi\frac{1}{\sqrt{1+K}},\nonumber\\
\lineup = c\left[(1+K)-(1+K)\s \frac{1}{1+K}
\sb(1+K)\right]Bc\frac{1}{1+K}.\label{eq:nonreal}
\end{eqnarray}
Now the square root factors do not appear, and the solution requires 
one fewer Schwinger integration. The non-real solution may 
be more appropriate for a potential generalization to the superstring, 
since at present we do not have a controlled solution for the superstring 
tachyon vacuum which satisfies the reality condition \cite{BerkVac}. 
The solution satisfies a linear gauge condition \cite{simple}:
\begin{equation}\mathcal{B}_{\frac{1}{\sqrt{1+K}},\frac{1}{\sqrt{1+K}}}\Psi = 
0.\label{eq:simp_gauge}\end{equation}
This is an example of a so-called {\it dressed Schnabl gauge}, 
$\mathcal{B}_{F,G}=0$, where the operator $\mathcal{B}_{F,G}$ is defined
\begin{equation}\mathcal{B}_{F,G} \equiv F\half\mathcal{B}^-\Big(F^{-1}[\ 
\cdot\ ]
G^{-1}\Big)G.\end{equation}
$\mathcal{B}^-$ is the BPZ odd component of Schnabl's $\mathcal{B}_0$ 
\cite{RZ} and $F,G$ are any pair of states in the wedge 
algebra.\footnote{To check the 
gauge condition, note that $\half\mathcal{B}^-$ is a derivation of the 
star product satisfying $\half\mathcal{B}^-K=B$ and it annihilates all 
other fields in the algebra.} The Schnabl gauge corresponds to the special 
case $\mathcal{B}_0=\mathcal{B}_{\sqrt{\Omega},\sqrt{\Omega}}=0$. We will 
discuss the analogous solution in Schnabl gauge in section \ref{sec:Schnabl}.

\section{Energy and Closed String Tadpole}

We now discuss two important gauge invariant quantities associated with
the solution: the spacetime action, and the so-called 
{\it Ellwood invariant} \cite{overlap1,overlap2,overlap3,overlap4,tadpole}, 
which is closely related to the closed string tadpole 
amplitude \cite{tadpole} and the boundary state \cite{KOZ,KMS}. Usually 
the computation of these quantities is a core technical obstacle 
for an analytic solution. But for us it will require very little work, since 
the computations almost immediately reduce to those of the tachyon vacuum, 
which are already described in \cite{simple}.

Let us start by computing the spacetime action:\footnote{We set the 
open string coupling constant to unity. This means that the disk 
partition function in $\BCFT_0$ must be normalized to the volume of the 
reference D-brane to compute the correct energy.}
\begin{equation}S = \Tr\left[-\frac{1}{2}\Psi Q\Psi -\frac{1}{3}\Psi^3\right],
\end{equation}
where we use $\Tr[\cdot]$ to denote the 1-string vertex (or Witten integral).
Since we consider time-independent configurations, really we are interested 
in the energy, which is minus the action divided by the volume of the time 
coordinate:
\begin{equation}E=-\frac{S}{\mathrm{Vol}(X^0)}.\end{equation}
For us, the volume of time must be infinite otherwise the timelike Wilson-line
alters the physical interpretation of the solution. Still we 
can compactify time and consider the limit when the volume goes to 
infinity. This has the effect of normalizing the disk partition 
function in the timelike component of the $X^0$ BCFT to unity (for the 
purposes of the energy computation). Plugging in the solution 
$\Psi=\Psi_\tv+\Phi$ we find:  
\begin{equation}E = \frac{1}{\mathrm{Vol}(X^0)}\left(-\frac{g_0}{2\pi^2}+
\Tr\left[-\frac{1}{2}\Phi Q_{\Psi_\tv}\Phi -\frac{1}{3}\Phi^3\right]\right),
\end{equation}
where the term $-\frac{g_0}{2\pi^2}$ comes from the energy of 
tachyon vacuum $\Psi_\tv$. Assuming the equations of motion, this simplifies 
to\footnote{The validity of the equations of motion contracted with the 
solution is notoriously subtle in string field theory 
\cite{Okawa,FKeom,twisted}. In the current context, one might worry about 
potentially ambiguous collisions of $\s$ and $\sb$. To clarify this question,
we considered a regularization of the solution $\Phi\to\Phi\Omega^\eps$. We 
found no evidence of problems in the $\eps\to 0$ limit. We confirmed 
this by explicit computation of the regularized kinetic and cubic terms of 
the action for the 2-brane solution \eq{double}, where both the four and 
six point functions of bcc operators are easily obtained in closed form.}
\begin{equation}
E = \frac{1}{\mathrm{Vol}(X^0)}\left(-\frac{g_0}{2\pi^2}+
\frac{1}{6}\Tr\left[\Phi^3\right]\right).
\end{equation}
Plugging in $\Phi = -\S \Psi_\tv \Sb$ and using $\Sb\S=1$, we find
\begin{equation}
\Tr[\Phi^3]=-\Tr\left[\Psi_\tv^3\right]_{\BCFT_*},
\end{equation}
where the subscript $\BCFT_*$ indicates that the whole boundary in the 
correlator has $\BCFT_*$ boundary conditions. Except for the normalization 
of the disk partition function, this is exactly the computation of the cubic 
term in the action for $\Psi_\tv$, and by standard manipulations we find
\begin{equation}E 
= \frac{1}{\mathrm{Vol}(X^0)}\left(-\frac{g_0}{2\pi^2}
+\frac{g_*}{2\pi^2}\right),
\end{equation}
which is the expected energy difference between $\BCFT_0$ and $\BCFT_*$.

Next we compute the Ellwood invariant \cite{tadpole},
\begin{equation}\Tr_\mathcal{V}[\Psi],\end{equation}
where $\Tr_\mathcal{V}[\cdot]$ is the 1-string vertex with a midpoint 
insertion of an on-shell closed string vertex operator of the form
$\mathcal{V}=c\overline{c}V^\mathrm{matter}$. Based on examples 
and general arguments \cite{tadpole,KOZ,KMS,Ishibashi}, the Ellwood invariant 
is believed to compute the shift in the closed string tadpole amplitude between
$\BCFT_0$ and $\BCFT_*$, or equivalently the shift in the on-shell part of 
the boundary state \cite{KMS}:
\begin{equation}\Tr_\mathcal{V}[\Psi] = \frac{1}{4\pi i}\Big(\langle 
\mathcal{V}|c_0^-|B_0\rangle -\langle\mathcal{V}|c_0^-|B_*\rangle\Big),
\label{eq:Ell_exp}
\end{equation}
where $|B_0\rangle$ is the boundary state in $\BCFT_0$ and $|B_*\rangle$ is 
the boundary state in $\BCFT_*$. The contribution from $\BCFT_0$ appears 
automatically from $\Psi_\tv$, and looking at the contribution from $\Phi$,
\begin{equation}\Tr_\mathcal{V}[\Phi]=-\Tr_\mathcal{V}[\S\Psi_\tv\Sb] 
= -\Tr_\mathcal{V}[\Psi_\tv]_{\BCFT_*}=-\frac{1}{4\pi i}\langle
\mathcal{V}|c_0^-|B_*\rangle,\label{eq:startad}\end{equation}
we get the contribution from $\BCFT_*$. Therefore the solution correctly 
describes the shift in the closed string tadpole amplitude.

More interesting than the closed string tadpole is the full 
BCFT boundary state. A rigorous approach would follow the construction of 
Kiermaier, Okawa, and Zwiebach \cite{KOS},  but for present purposes it 
is enough to take the more pragmatic route of \cite{KMS}, which requires 
little more than the above computation of the Ellwood invariant. The key 
observation of \cite{KMS} is that we can compute the overlap of the boundary 
state with any matter primary $V$ provided we tensor with an auxiliary BCFT 
(with vanishing central charge) and compute the Ellwood invariant with a 
modified matter vertex operator
\begin{equation}w V,\end{equation}
where $w$ lives in the auxiliary BCFT and cancels the conformal 
weight of $V$ so that the combination is a weight $(1,1)$ 
matter+auxiliary primary. From the form of the solution, it is clear that 
the boundary state of the solution $|B_\Psi\rangle$ can be factorized into 
timelike/spacelike components:\footnote{Note that disk 1-point 
functions in $\BCFT_*$ represent the contraction of a bulk vertex operator
with a closed string state which is manifestly space/time factorized on 
account of the factorization of $\s$ and $\sb$. The ghost factor of 
the boundary state is universal in bosonic 
string theory \cite{KMS}.} 
\begin{equation}|B_\Psi\rangle = |B_\Psi\rangle^{X^0}\otimes 
|B_\Psi\rangle^{c=25}\otimes|B\rangle^{bc}.
\end{equation}
The matter part of the boundary state can be expressed as a sum of Virasoro
Ishibashi states $|V_\alpha\rangle\!\rangle$ associated with spinless primaries
$V_\alpha$ in the time/spacelike sectors, with appropriate coefficients:
\begin{equation}|B_\Psi\rangle = \left(\sum_{{\alpha =\atop 
X^0\ \mathrm{primaries}}}
 n^{\alpha}_\Psi 
|V_{\alpha}\rangle\!\rangle\right)\otimes 
\left(\sum_{{\beta =\atop c=25\ \mathrm{primaries}}} n^\beta_\Psi 
|V_\beta\rangle\!\rangle\right)
\otimes|B\rangle^{bc}.\end{equation}
The coefficients $n_\Psi^\alpha$ represent disk one-point functions of 
$V^\alpha$ with the appropriate boundary condition. Here $V^\alpha$ is the 
dual vertex operator to $V_\alpha$, so that 
$\langle V^\alpha|V_\beta\rangle=\delta^\alpha_\beta$. Following the 
proposal of \cite{KMS} we can compute these coefficients with the Ellwood 
invariant:\footnote{A potentially subtle point in this approach is that it 
requires a definition of the solution in an enhanced BCFT which includes 
the auxiliary factor. For the solution \eq{sol}, this only requires taking 
$K\to K+K^\mathrm{aux}$, where $K^\mathrm{aux}$ represents an insertion of the 
energy-momentum tensor in the auxiliary BCFT.}
\begin{equation}n_\Psi^\alpha = 2\pi i\Tr_{\mathcal{V}^\alpha}[\Phi],\ \ \ \ \ 
\mathcal{V}^\alpha = c\overline{c} (w^\alpha V^\alpha).\end{equation}
From \eq{startad} it is clear that this computes the disk one-point function
in $\BCFT_*$ (provided $\langle w^\alpha\rangle_\mathrm{disk}=1$), and 
therefore we recover the expected boundary state. One important point, which 
we can now address in a more explicit manner, is the extent to which the 
timelike Wilson line effects the physical interpretation of the solution. For
this we need to investigate the coefficients $n_\Psi^\alpha$ for the timelike 
factor of the boundary state. Evaluating the vacuum correlator for the 
spacelike components and mapping to the upper half plane, we find
\begin{equation}n_\Psi^\alpha = 
\mathrm{const.}\times\left\langle \exp\left[\sqrt{h}\int_{-\infty}^\infty
ds\, i\d X^0(s)\right]V^\alpha(i,\overline{i})\right
\rangle_{\mathrm{UHP}}^{X^0,\BCFT_0},
\end{equation}
where, by evaluating the correlator in the timelike component of 
$\BCFT_0$, we bring out the Wilson line boundary interaction with coupling 
$\sqrt{h}$ given by the conformal weight $h$ of the spacelike bcc operators. 
Since $i\d X^0$ is a chiral operator, we can regularize the boundary 
interaction by simply deforming the contours away from the boundary 
\cite{Schomerus}. Deforming the contours to surround the bulk insertion, 
we potentially pick up a residue from a pole in the OPE if $i\d X^0$ with 
$V^\alpha$. The only primary with such a pole is a plane wave
\begin{eqnarray}
i\d X^0(z) e^{ik X^0}(w,\overline{w}) \sim \frac{k}{2}
\frac{1}{z-w}e^{i kX^0},\ \ \ \ z\to w,
\end{eqnarray}
but momentum conservation in the 1-point function forces $k=0$. Therefore 
the $i\d X^0$ contours close without hitting a pole, and we find:
\begin{equation}
n_\Psi^\alpha =\mathrm{const}\times\langle V^\alpha(0,0)
\rangle_{\mathrm{UHP}}^{X^0,\BCFT_0}.
\end{equation}
The timelike component of the boundary state is unchanged by the solution.
This is consistent with the expectation that the timelike Wilson line 
is pure gauge.

\section{Cohomology and Background Independence}
\label{sec:coh}

The physical excitations of the solution $\Psi$ are described by the 
cohomology of the shifted kinetic operator,
\begin{equation}Q_\Psi = Q+[\Psi,\cdot].\end{equation}
This cohomology should be the same as the cohomology of $Q$ in $\BCFT_*$. 
With a few qualifications, we will show that this is indeed the case.

Let us start by assuming that the solution $\Psi$ is a marginal deformation, 
since here the argument is uncomplicated. In this case the disk partition 
functions are equal $g_0=g_*$ and $\S$ and $\Sb$ are inverses in both 
directions:
\begin{equation}\S\Sb=\Sb\S = 1,\ \ \ \ (\mathrm{marginal\ case}).
\end{equation}
Note also the relations\footnote{We define
$Q_{\Phi_1\Phi_2} A \equiv QA + \Phi_1 A+(-1)^A A\Phi_1$. This is the kinetic
operator for a stretched string between classical solutions $\Phi_1$ and 
$\Phi_2$.}
\begin{equation}Q_{\Psi0}\Sigma = 0;\ \ \ \ Q_{0\Psi}\Sb = 0.\label{eq:QSSb}
\end{equation} 
For example the first can be demonstrated as follows: 
\begin{eqnarray}
Q_{\Psi 0}\S \lineup = Q\S + \Psi \S,\nonumber\\
\lineup = Q\S +\Psi_\tv\S - \S \Psi_\tv \Sb \S,\nonumber\\
\lineup = Q\S +\Psi_\tv\S - \S \Psi_\tv,\nonumber\\
\lineup = Q_{\Psi_\tv}\S =0.
\end{eqnarray}
and the second follows similarly. With these ingredients, we can define 
an isomorphism between states in $\BCFT_0$ and $\BCFT_*$:
\begin{eqnarray}
\varphi_0\lineup = \ f(\varphi_*)\ \,=\S\varphi_*\Sb;\label{eq:f}\\
\varphi_*\lineup = f^{-1}(\varphi_0)=\Sb\varphi_0\S,\label{eq:finv}
\end{eqnarray}
satisfying 
\begin{equation}f\circ f^{-1} = 1_{\BCFT_0};\ \ \ \ \ \ f^{-1}\circ f = 
1_{\BCFT_*}.\label{eq:iso}\end{equation}
where $\varphi_0$ and $\varphi_*$ are suitably well-behaved states in $\BCFT_0$
and $\BCFT_*$, respectively. Furthermore, \eq{QSSb} implies that $f$ and 
$f^{-1}$ satisfy
\begin{equation}Q_\Psi f(\varphi_*) = f(Q\varphi_*);\ \ \ \ \ \ 
Qf^{-1}(\varphi_0) = f^{-1}(Q_\Psi\varphi_0),\label{eq:Qf}\end{equation}
so we have an isomorphism of cohomologies. In summary, if 
$\Psi$ is a marginal deformation, the cohomology of $Q_\Psi$ in $\BCFT_0$ is 
identical to the cohomology of $Q$ in $\BCFT_*$. 

The non-marginal case is more subtle. Here, we still have equations 
\eq{QSSb} and \eq{Qf}, but since $\S\Sb\neq 1$ equation \eq{iso} 
is replaced with 
\begin{equation}f\circ f^{-1} = \left(\frac{g_*}{g_0}\right)^2
\times 1_{\BCFT_0};
\ \ \ \ \ \ f^{-1}\circ f = 1_{\BCFT_*}.\label{eq:noniso0}\end{equation}
Thus composition of $f$ and $f^{-1}$ would seem to be non-associative, and it 
is not clear that we have a well-defined isomorphism. This is an 
indication that we need to be more careful about domains. To start, let 
$\mathcal{H}_0$ denote the state space of $\BCFT_0$ and $\mathcal{H}_*$ 
the state space of $\BCFT_*$. Consider a subspace of ``perturbative'' 
states in $\BCFT_*$,
\begin{equation}
\mathcal{H}_*^{\mathrm{pert}}\subset\mathcal{H}_*.
\end{equation}
Loosely speaking, $\mathcal{H}_*^{\mathrm{pert}}$ consists of string fields 
which produce no collisions with bcc operators upon multiplication with 
$\S$ and $\Sb$. This includes, for example, perturbative Fock states. Mapping
$\mathcal{H}_*^{\mathrm{pert}}$ using $f$ in \eq{f} defines a subspace of 
states in $\BCFT_0$:
\begin{equation}f\circ\mathcal{H}_*^{\mathrm{pert}}\subset\mathcal{H}_0.
\end{equation}
Using the inverse map $f^{-1}$ in \eq{finv}, it is clear that this subspace 
in $\BCFT_0$ is isomorphic to the subspace $\mathcal{H}_*^{\mathrm{pert}}$ 
in $\BCFT_*$. Therefore, if we look for the cohomology of $Q_\Psi$ within 
$f\circ\mathcal{H}_*^{\mathrm{pert}}$, it will be the same as the
cohomology of $Q$ in $\mathcal{H}_*^{\mathrm{pert}}$. 

This may not appear to be fully satisfactory. While 
$\mathcal{H}_*^{\mathrm{pert}}$ represents a fairly generic class of 
states in $\BCFT_*$, $f\circ\mathcal{H}_*^{\mathrm{pert}}$ are 
rather peculiar states in $\BCFT_0$. Our main reason for restricting domains 
is to have a well-defined isomorphism between the state spaces, but this is 
probably more than is needed to prove the isomorphism of cohomologies. 
Let us explain this with a degenerate example. Consider the tachyon vacuum, 
where $\S=\Sb=0$. In this case, $f\circ\mathcal{H}_*^{\mathrm{pert}}$ is the 
zero vector, which of course is consistent with the absence of cohomology. 
But the tachyon vacuum kinetic operator has trivial cohomology not just when 
computed on the zero vector, but also when computed for fairly arbitrary 
states in $\BCFT_0$.\footnote{For Schnabl's solution and related solutions, 
the absence of cohomology is clear for reasonably well-behaved states 
\cite{cohomology}. However, this remains a subtle question. There 
are indications that cohomology at exotic ghost numbers appears for the 
Siegel gauge condensate \cite{Imbimbo}, and for the identity-based tachyon 
vacuum solution of Takahashi and Tanimoto \cite{TT,TT2,TTcoh}.} In a similar
way, for general backgrounds the cohomology of $Q_\Psi$ may be correct 
even when computed outside $f\circ\mathcal{H}_*^{\mathrm{pert}}$, though at 
present we will not attempt to make this statement precise.

Let us point out an important consequence of our construction. Consider 
the action expanded around the solution $\Psi$:
\begin{equation}S = \frac{g_0-g_*}{2\pi^2}+\Tr\left[-\frac{1}{2}\Phi_0 Q_\Psi
\Phi_0-\frac{1}{3}\Phi_0^3\right].\end{equation}
Setting $\Phi_0=\S\Phi_*\Sb$ this becomes
\begin{equation}S = \frac{g_0-g_*}{2\pi^2}+\Tr\left[-\frac{1}{2}\Phi_* Q
\Phi_*-\frac{1}{3}(\Phi_*)^3\right]_{\BCFT_*}.\end{equation}
Thus we have recovered the string field theory formulated around 
$\BCFT_*$. This gives an astonishingly simple proof of background 
independence in open string field theory.\footnote{Previous analysis of 
this problem can be found in \cite{SenB1,SenB2,SenB3,SenB4,Ellwood}.}

\section{Schnabl gauge Solution}
\label{sec:Schnabl}

The solution we have been working with is simple, but it is close 
to being singular from the perspective of the identity string field 
\cite{IdSing} and for some purposes it may be necessary to work with a 
more regular solution. Ideally, we would like to find an analogue of 
\eq{sol} in Schnabl gauge \cite{Schnabl}:
\begin{equation}\mathcal{B}_0\Psi_\Sch=0.\end{equation}
The expectation is that in Schnabl gauge the solution should be built from 
wedge states $\Omega^\alpha$ with $\alpha$ strictly greater than one. 
By contrast, the original solution \eq{sol} is built from wedge states 
all the way down to the identity string field.

There is a simple transformation relating solutions in different dressed 
Schnabl gauges:
\begin{equation}
\Psi_{F} = \sqrt{F/f}\,\Psi_{f}
\frac{1}{1+B\,\frac{1-F/f}{K}\,\Psi_{f}}\sqrt{F/f},
\end{equation}
where $\Psi_{f}$ is a solution in $\mathcal{B}_{\sqrt{f},\sqrt{f}}$-gauge and 
$\Psi_{F}$ is a solution in $\mathcal{B}_{\sqrt{F},\sqrt{F}}$-gauge. This is a 
version of the Zeze map, introduced in \cite{currents}. In the current 
situation, we want to map from a solution satisfying 
$\mathcal{B}_{\frac{1}{\sqrt{1+K}},\frac{1}{\sqrt{1+K}}}=0$ to Schnabl gauge
$\mathcal{B}_0=\mathcal{B}_{\sqrt{\Omega},\sqrt{\Omega}}=0$, and the 
transformation becomes\footnote{This formula actually takes the same form 
in transforming $\mathcal{B}_{\frac{1}{\sqrt{1+K}},\frac{1}{\sqrt{1+K}}}$ gauge
to any $\mathcal{B}_{\sqrt{F},\sqrt{F}}$ gauge, with the replacement 
$\Omega\to F$.}
\begin{equation}
\Psi_\Sch = \sqrt{\Omega(1+K)}\Psi
\frac{1}{1+B\Delta\Psi}\sqrt{(1+K)\Omega},\label{eq:Sch1}
\end{equation}
where $\Delta$ is the string field 
\begin{equation}\Delta \equiv \frac{1-\Omega}{K}-\Omega.\end{equation}
The field $\Delta$ has a special interpretation. Given any 
$\mathcal{B}_{\sqrt{F},\sqrt{F}}$-gauge there are two distinguished elements 
of the wedge algebra: the ``security strip'' $F$, which surrounds the 
operator insertions in the solution, and the ``homotopy field'' 
$\frac{1-F}{K}$, which appears (for example) in the homotopy operator 
which trivializes the cohomology around the tachyon vacuum 
\cite{cohomology,SSFII}. The simplest gauge 
$\mathcal{B}_{\frac{1}{\sqrt{1+K}},\frac{1}{\sqrt{1+K}}}=0$
has the special property that the security strip and 
homotopy field are equal. Therefore, the field $\Delta$ characterizes the 
failure of Schnabl gauge to be ``simple.'' Substituting \eq{sol} we find 
a more explicit expression for the Schnabl-gauge solution:
\begin{equation}
\Psi_\Sch = \sqrt{\Omega}c \frac{1}{1+\left(1-(1+K)
\displaystyle{\s\frac{1}{1+K}\sb}\right)
\Delta}\left(1-(1+K)\s\frac{1}{1+K}\sb\right)(1+K)Bc\sqrt{\Omega}.
\label{eq:Sch2}
\end{equation}
We would like to define this as a power series in $\Delta$. 
Computing the $\Delta^n$ correction in the Fock space requires knowledge 
of a $2n+3$-point correlator with a test state in $\BCFT_0$ and 
$2n+2$ bcc operators. Such correlators would be difficult to compute 
in general, and the original solution \eq{sol} is certainly simpler. 

One immediate question is whether a power series in $\Delta$ converges. 
We do not know the answer to this question in general, but
if our goal is to regulate the identity-like nature of the solution 
\eq{sol}, we can choose any number of dressed Schnabl gauges where the 
analogue of $\Delta$ can be taken to be as small as we like, and presumably 
the power series can be made to converge. Still, 
in the case of Schnabl gauge we can get some insight into the nature of 
convergence by looking at the the case $\s=\sb=0$. This gives Schnabl's 
solution for the tachyon vacuum, expressed in the form
\begin{equation}\Psi_\Sch = \sqrt{\Omega}c\frac{1+K}{1+\Delta}Bc\sqrt{\Omega}.
\end{equation}  
Actually, for illustrative purposes we can ignore the ghosts and look at the 
ghost number zero toy model \cite{Schnabl}:
\begin{equation}\frac{1+K}{1+\Delta}\Omega.\end{equation}
To see convergence in powers of $\Delta$, consider the coefficient 
of $L_{-2}|0\rangle$, which can be computed by the formula 
\cite{Schnabl,exotic}
\begin{eqnarray}L_{-2}|0\rangle\ \mathrm{coefficient}\lineup 
=-\frac{1}{3}+\frac{4}{3}
\int_0^\infty dK\ K e^{-K} \left(\frac{1+K}{1+\Delta}\Omega\right),\nonumber\\
\lineup =-\frac{1}{3}+\frac{4}{3}
\int_0^\infty dK\ K e^{-K} \left(\frac{K\Omega}{1-\Omega}\right),\nonumber\\
\lineup = -3+\frac{8}{3}\zeta(3).
\end{eqnarray}
Now expand this in powers of $\Delta$:
\begin{equation}
L_{-2}|0\rangle\ \mathrm{coefficient} 
=-\frac{1}{3}+\frac{4}{3}\sum_{n=0}^\infty(-1)^n
\int_0^\infty dK\ K(1+K) e^{-2K} \Delta^n.
\end{equation}
Using the method of steepest descent, the $n$th contribution to this sum for 
large $n$ can be estimated as 
\begin{equation}
(-1)^n\int_0^\infty dK\ K(1+K) e^{-2K} \Delta^n=
\sqrt{\frac{2\pi\Delta(\gamma)}{n |\Delta''(\gamma)|}}\Delta(\gamma)^n 
\gamma(1+\gamma)e^{-2\gamma}+\,.\,.\,.\,,
\end{equation}
where  $\Delta(\gamma)\approx 0.298426$ is the maximum value of $\Delta$ as 
a function of $K$, $\Delta''(\gamma)\approx -0.0736153$ is the second 
derivative of $\Delta$ at its maximum, and $\gamma\approx 1.79328$. Thus the 
$n$th term in the expansion in powers of $\Delta$ is exponentially suppressed, 
and the series converges fairly quickly. In fact, convergence is much faster 
than standard definition of Schnabl's solution as a power series in $\Omega$:
\begin{equation}
L_{-2}|0\rangle\ \mathrm{coefficient} 
=-\frac{1}{3}+\frac{4}{3}\sum_{n=0}^\infty
\int_0^\infty dK\ K^2 e^{-(2+n)K}.
\end{equation}
where the $n$th term contributes as $2/n^3$.

The power series expansion in $\Delta$ has another interesting 
property: It gives a definition of Schnabl's solution without the phantom 
term.\footnote{For some relevant discussions of the phantom term in Schnabl's
solution and other solutions, see 
\cite{Schnabl,simple,Integra,Okawa,FKeom,SSFII,phantom}.} To see that this 
is the case, compute the Ellwood invariant:
\begin{equation}\Tr_\mathcal{V}[\Psi_\Sch]
=\Tr_\mathcal{V}\left[c\frac{1+K}{1+\Delta}Bc\Omega\right]
=\sum_{n=0}^\infty(-1)^n
\Tr_\mathcal{V}\left[c(1+K)\Delta^n Bc\Omega\right].
\end{equation}
Using the well-known formula \cite{MS}
\begin{equation}\Tr_\mathcal{V}[c F Bc G]=-F(0)G'(0) 
\frac{1}{4\pi i}\langle \mathcal{V}|c_0^-|B_0\rangle,
\end{equation}
for $F,G$ states in the wedge algebra, we find 
\begin{equation}\Tr_\mathcal{V}[\Psi_\Sch]
=\sum_{n=0}^\infty(-1)^n \Delta(0)^n
\frac{1}{4\pi i}\langle \mathcal{V}|B_0\rangle.
\end{equation}
Since $\Delta(0)=0$, the sole contribution to the Ellwood invariant comes 
from the zeroth order term in the power series in $\Delta$. Dropping a 
BRST trivial piece, this is simply the zero momentum tachyon:
\begin{equation}\sqrt{\Omega}c\sqrt{\Omega}=\frac{2}{\pi}c_1|0\rangle.
\end{equation}
The source for the closed string does not come from a sliver-like phantom 
term.

\section{Tachyon Lump}
\label{sec:lumps}

Having completed the general discussion of the solution, we turn our 
attention to a specific (and fundamental) example: the tachyon lump, 
describing the formation of a D$(p-1)$-brane in the string field theory of a 
D$p$-brane. Previous numerical constructions of the tachyon lump in Siegel 
gauge are discussed in \cite{KMS,MSZ,Moeller_lump,Beccaria}. 
We will describe formation of the lump along a direction 
$X^1$ which has been compactified on a circle of radius $R$. The bcc 
operators describing this background are the Neumann-Dirichlet twist operators 
$\s_\ND,\sb_\ND$ of weight $\frac{1}{16}$, described for example in 
\cite{Dixon,Narain,Frohlich}. Tensoring with a Wilson line gives: 
\begin{equation}\s(s) = \s_\ND e^{iX^0/4}(s);\ \ \ \ \sb(s)=\sb_\ND 
e^{-iX^0/4}(s).\label{eq:lumpbcc}
\end{equation}
For our computations, the most important piece of information we need to know 
about the Neumann-Dirichlet twist operators is the three-point function 
with a plane wave, computed in \cite{Mukhopadhyay}:
\begin{equation}
\langle e^{inX^1/R}(s_1) \s_\ND(s_2)\sb_\ND(s_3)\rangle_{\mathrm{UHP}}^{X^1}
=\frac{2\pi 2^{-2\frac{n^2}{R^2}}e^{\frac{ina}{R}}}{|s_{12}s_{13}|^{
\frac{n^2}{R^2}}|s_{23}|^{\frac{1}{8}-\frac{n^2}{R^2}}},\label{eq:3pt}
\end{equation}
where the Dirichlet boundary condition is fixed to a position $a$ along the 
circle and $s_{ij}=s_i-s_j$.\footnote{When $X^1$ has Neumann boundary 
conditions, we will normalize the disk partition function in the $X^1$ 
BCFT to the spacetime volume $2\pi R$. This means that when $X^1$ has 
Dirichlet boundary conditions, the disk partition function must be 
normalized to $2\pi$ to obtain the correct ratio of tensions. This is the 
origin of the factor of $2\pi$ in \eq{3pt}.}

In light of earlier discussion, one immediate question about 
the proposed lump solution is why it does not represent a marginal 
deformation. To understand this we must determine the fate of the 
marginal operator \eq{V}:
\begin{equation}V = \s\d \sb.\end{equation}
Finding $V$ requires knowledge of the subleading structure of the 
$\s_\ND$-$\sb_\ND$ OPE. The leading term is proportional to the identity 
operator, and the next to leading term must be proportional to the first
cosine harmonic on the circle. The precise coefficients can be derived 
from the 3-point function \eq{3pt}:\footnote{The OPEs \eq{NDOPE}, \eq{ssbOPE},
and \eq{sdsbOPE} are correct for $R>1/\sqrt{2}$, otherwise the contribution 
from the first cosine harmonic is subleading to descendents of the identity 
and $i\d X^0$. In addition, \eq{ssbOPE} and \eq{sdsbOPE} assume $R<\sqrt{2}$ 
otherwise the contribution from $i\d X^0$ is subleading to second or higher 
cosine harmonics.}
\begin{equation}\s_\ND (s)\sb_\ND(0)=\frac{1}{s^{1/8}}\cdot\frac{1}{R}\,
+\,\frac{1}{s^{1/8-1/R^2}}\cdot
\frac{2^{-2/R^2+1}}{R}\cos\left(\frac{X^1-a}{R}\right)(0)+
\ .\,.\,.\, ,\ \ \ (s>0).\label{eq:NDOPE}
\end{equation}
The bcc operators used in the solution must therefore have the 
OPE
\begin{equation}
\s(s)\sb(0) = \frac{1}{R}
+s^{1/R^2}\cdot\frac{2^{-2/R^2+1}}{R}\cos\left(\frac{X^1-a}{R}
\right)(0)+s\cdot \frac{i}{4}\d X^0(0)+\ .\,.\,.\, ,\ \ \ (s>0).
\label{eq:ssbOPE}
\end{equation}
The marginal operator is obtained by taking the derivative 
with respect to $s$ and considering the limit $s\to 0$:
\begin{equation}
\s(s)\d \sb(0) = s^{1/R^2-1}\cdot\frac{-2^{-2/R^2+1}}{R^3}
\cos\left(\frac{X^1-a}{R}\right)(0)
-\frac{i}{4}\d X^0(0)+\ .\,.\,.\, ,\ \ \ \ (s>0).\label{eq:sdsbOPE}
\end{equation}
The fate of the $s\to 0$ limit depends on the compactification radius $R$:
\begin{itemize}
\item $R>1$ (Relevant deformation): The ``marginal operator'' is infinite, 
or more specifically, a divergent constant times the relevant matter operator 
$\cos\left(\frac{X^1-a}{R}\right)$. Since the marginal operator does 
not exist, there is obviously no corresponding family of conformal 
boundary conditions connecting $\BCFT_0$ and $\BCFT_*$.
\item $R=1$ (Marginal deformation): In this case we have a marginal operator
\begin{equation}V = -\frac{1}{2}\left[\cos(X^1-a)
+\frac{i}{2}\d X^0\right].\end{equation}
Since this operator has regular self-OPE, it can be used to construct
a solution for nonsingular marginal deformations in Schnabl gauge 
\cite{KROZ,Sch_marg} or following KOS \cite{KOS}. Ignoring the
timelike Wilson line, this operator obviously represents the cosine marginal 
deformation on the circle at self-dual radius \cite{Ludwig}. In our 
conventions, the moduli space of the cosine deformation 
$\lambda\cos(X^1-a)$ is periodic with the identification
$\lambda \sim \lambda+1$. $\lambda=1/2$ represents the critical value 
where the boundary condition becomes Dirichlet, which is why an overall 
factor of $1/2$ appears in $V$.
\item $R<1$ (Irrelevant deformation): In this case we have the marginal 
operator
\begin{equation}V=-\frac{i}{4}\d X^0.\end{equation}
This operator turns on a timelike Wilson line, but all information about 
the formation of the D$(p-1)$-brane has been lost. In this case, the solution
is more naturally understood in the $T$-dual picture $R\to 1/R$, where it 
represents the reverse process of formation of a higher dimensional 
D$p$-brane in the string field theory on a D$(p-1)$-brane. While we are able
to construct a marginal operator, because it has singular self-OPE the 
state $e^{-(K+V)}$ assumed to exist in \eq{def} is divergent. We could 
renormalize the boundary interaction to create the Wilson line, but the
formal argument of \eq{farg} connecting $\BCFT_0$ to $\BCFT_*$ will no 
longer apply. 
\end{itemize}
Therefore, the fact that the tachyon lump is not a marginal deformation 
does not pose a contradiction for the solution. As a consistency check on 
this picture, we verify the $\s$-$\sb$ OPE using the four-point 
function of Neumann-Dirichlet twist fields in appendix \ref{app:twist}.

One important thing to compute from the solution is the position space 
profile of the tachyon field. This gives a concrete (but gauge dependent) 
spacetime picture of the solitonic lump describing the lower dimensional
D-brane. To construct the tachyon profile, we expand the solution in the 
Fock space basis and focus on the tachyon state $|T\rangle$, which can be 
further expanded in plane wave harmonics on the circle
\begin{equation}
|T\rangle=\sum_{n\in \mathbb{Z}} t_n|T_n\rangle,\ \ \ \ \ |T_n\rangle\equiv
c\,e^{i n X^1/R}(0)|0\rangle,
\end{equation}
where $t_n$ are Fourier coefficients. The tachyon profile is defined by the 
function
\begin{equation}
t(x)=\sum_{n\in\mathbb{Z}} t_n\, e^{i \frac{nx}{R}}.
\end{equation}
Define a ``test state'' $|\tilde{T}_n\rangle$ dual to the $n$th tachyon 
harmonic:\footnote{For the purposes of this computation, we will compactify 
all directions besides $X^1$ on circles of unit circumference, so that the 
norm of the $SL(2,\mathbb{R})$ vacuum is $2\pi R$. Strictly speaking, the 
time direction is noncompact so the vacuum should be delta function normalized.
Then the dual test state $|\tilde{T}_n\rangle$ should include a superposition 
of plane waves in the time direction which creates an eigenstate of the zero 
mode of $X^0$. The tachyon coefficients computed in this way turn out to be 
the same as when compactifying time.}
\begin{equation}|\tilde{T}_n\rangle = 
-\frac{1}{2\pi R}\, c\d c\,e^{-in X^1/R}(0)|0\rangle.\end{equation}
By construction, this satisfies $\langle \tilde{T}_m,T_n\rangle=\delta_{mn}$. 
The tachyon coefficients $t_n$ can be computed from the contraction
\begin{eqnarray}
t_n \lineup = \langle \tilde{T}_n,\Psi\rangle,\nonumber\\
\lineup =-\frac{1}{2\pi R}\left(\frac{2}{\pi}\right)^{-1+n^2/R^2}
\Tr\Big[\sqrt{\Omega}(c\d c e^{-in X^1/R})\sqrt{\Omega}\Psi\Big].
\end{eqnarray}
In the last step we mapped the test vertex operator to the sliver coordinate 
frame (using $f_\mathcal{S}(z)=\frac{2}{\pi}\tan^{-1}(z)$ \cite{Schnabl,RZO})
where we can compute the contraction as a correlation function on the 
cylinder. To have simpler formulas we will use the non-real form on the 
solution \eq{nonreal}, which eliminates the square roots and places the 
``security strip'' $\frac{1}{1+K}$ completely to the right of 
operator insertions. Furthermore, it is convenient to rewrite the solution 
in a form which isolates the zero momentum contribution:
\begin{equation}
\Psi = \left(\frac{R-1}{R}\,c(1+K)Bc\frac{1}{1+K}\right) 
- \left(c\d\s \frac{B}{1+K}\sb (1+K)c\frac{1}{1+K}\right).
\label{eq:Psilump}\end{equation}
We recover the solution expressed in \eq{sol} by substituting 
$\d\s = [1+K,\s]$ and using $\s\sb=\frac{1}{R}$. The first term gives
the sole contribution at zero momentum, and is proportional the the ``simple'' 
tachyon vacuum, while the second term is the KOS solution \eq{KOS}. Note that
when the lump is marginal at $R=1$, the tachyon vacuum term disappears 
and we are left with the KOS solution, as expected. Since the tachyon 
coefficient of the ``simple'' tachyon vacuum was already computed 
in \cite{simple}, we immediately obtain
\begin{equation}t_0 \approx \frac{R-1}{R}\times 0.2844.
\label{eq:t0}\end{equation}
The remaining tachyon coefficients come from the KOS solution. In \cite{KOS},
KOS gave the general form of the contraction of the solution with 
any state of the form $\phi = -c\d c \phi^\mathrm{m}$, with $\phi^\mathrm{m}$
a matter primary of weight $h$: 
\begin{equation}\langle \phi,\Psi\rangle = C_\phi\times g(h).\end{equation}
Here $C_\phi$ is the 3-point function of $\phi^\mathrm{m}$ with the
two bcc operators:
\begin{equation}C_\phi\equiv \langle \sb(\infty)\phi^\mathrm{m}(1)\s(0)
\rangle^\mathrm{matter}_{\mathrm{UHP}},
\end{equation}
and $g(h)$ is a universal function which depends only on the weight $h$ of 
$\phi^m$, and not on the details of the boundary conformal field theories 
in question. For us, the function $g(h)$ takes a somewhat different form than
originally written by KOS since we use the non-real solution, and 
most importantly, the formulas written in \cite{KOS} assume the existence 
of a marginal operator $V=\s\d \sb$ which turns out to be divergent for 
relevant deformations. After some computation we find\footnote{The two 
terms come from further reexpressing the KOS solution in the form 
\begin{equation}\Psi_\mathrm{KOS} = -c\d \s \frac{B}{1+K}c\sb 
-c\d \s \frac{B}{1+K}\sb\d c\frac{1}{1+K}\end{equation}}
\begin{equation}g(h) = g_1(h)+g_2(h),\end{equation}
where
\begin{eqnarray}
g_1(h) \lineup = -h\! \int_0^\infty ds\,e^{-s}
\left(\frac{4}{L}\cot\theta_{\frac{1}{2}}\right)^{h-1}
\left(\frac{1}{L}-\frac{1}{\pi}\sin 2\theta_s\right),\ \ \ \ \ \ (L=s+1);
\label{eq:g1}\\
g_2(h) \lineup = 2h\! \int_0^\infty \!\!\!ds\!\!
\int_{1/2}^\infty \!dy
\frac{e^{-L+1}\sin \theta_{s+\frac{1}{2}}}{L\sin^2\theta_\frac{1}{2}}
\left(\frac{2\sin\theta_s}{L\sin\theta_\frac{1}{2}
\sin\theta_{s+\frac{1}{2}}}\right)^{h-1}\!\!\!(\theta_\frac{1}{2}
\cos\theta_{s+\frac{1}{2}}-\cos\theta_s\sin\theta_\frac{1}{2}),\nonumber\\
\lineup\ \ \ \ \ \ \ \ \ \ \ \ \ \ \ \ \ \ \ \ \ \ \ \ \ \ \ \ \ 
\ \ \ \ \ \ \ \ \ \ \ \ \ \ \ \ \ \ \ \ \ \ \ \ \ \ \ \ \ 
\ \ \ \ \ \ \ \ \ \ \ \ \ \ \ \ \ \ \
(L=\half+s+y).\label{eq:g2}
\end{eqnarray}
The angular parameters in these integrals are defined
\begin{equation}\theta_\ell\equiv \frac{\pi \ell}{L},\end{equation}
where $L$ appears in parentheses accompanying the respective integral. 
Therefore when $n\neq 0$ the tachyon coefficients can be computed
\begin{equation}t_n = \frac{2^{-2n^2/R^2}}{R} g(n^2/R^2),\end{equation}
where we substituted
\begin{equation}C_{\tilde{T}_n} = 
\left\langle \sb(\infty)\frac{1}{2\pi R}e^{-in X^1/R}(1)\s(0)
\right\rangle^\mathrm{matter}_\mathrm{UHP}
=\frac{2^{-2n^2/R^2}}{R},
\end{equation}
which follows immediately from \eq{3pt}. We center the lump at the origin 
by taking the constant $a$ in \eq{3pt} to be zero. With this definition, 
the tachyon coefficients satisfy $t_{-n}=t_n$.

\begin{figure}
\begin{center}
\resizebox{2.3in}{1.4in}{\includegraphics{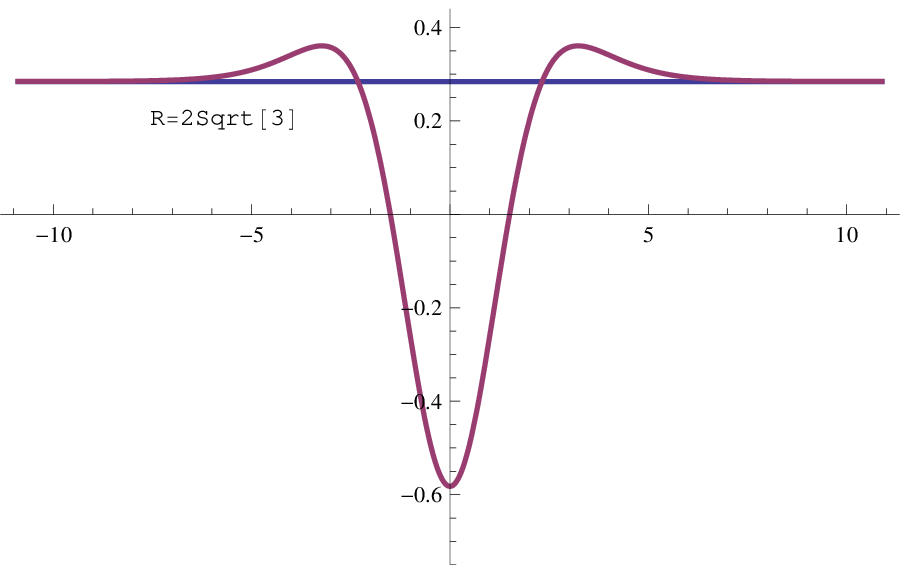}}\ \ \ \ \ \ \ \ \ \ 
\resizebox{2.3in}{1.4in}{\includegraphics{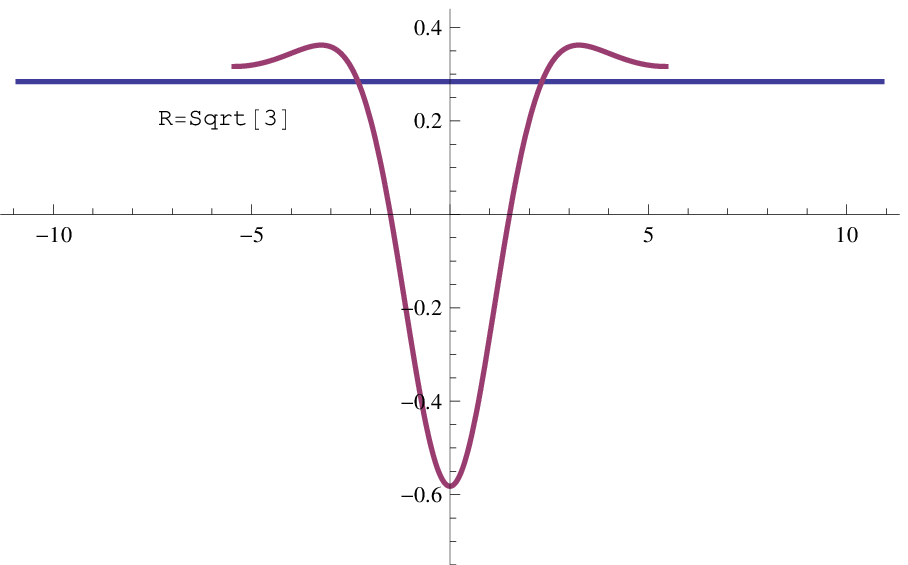}}\\ \  \\
\resizebox{2.3in}{1.4in}{\includegraphics{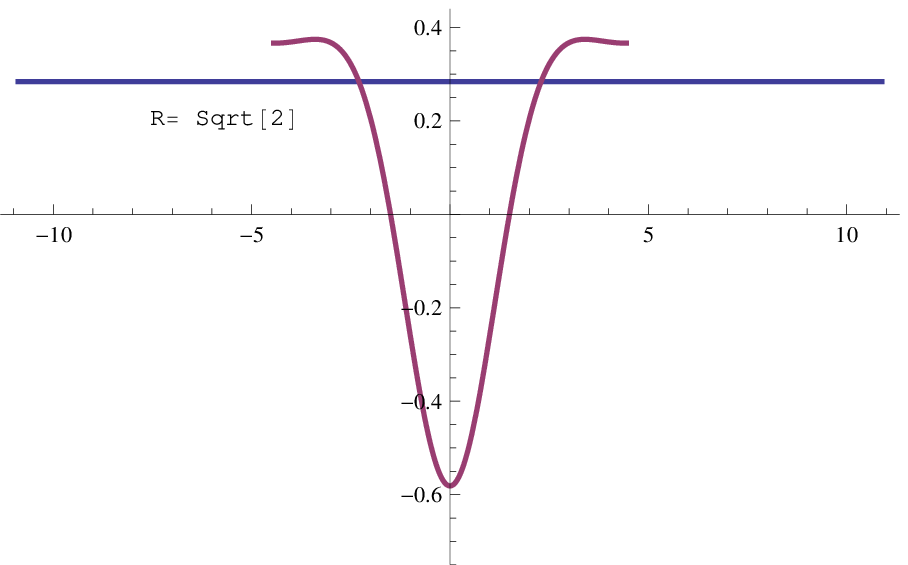}}\ \ \ \ \ \ \ \ \ \ 
\resizebox{2.3in}{1.4in}{\includegraphics{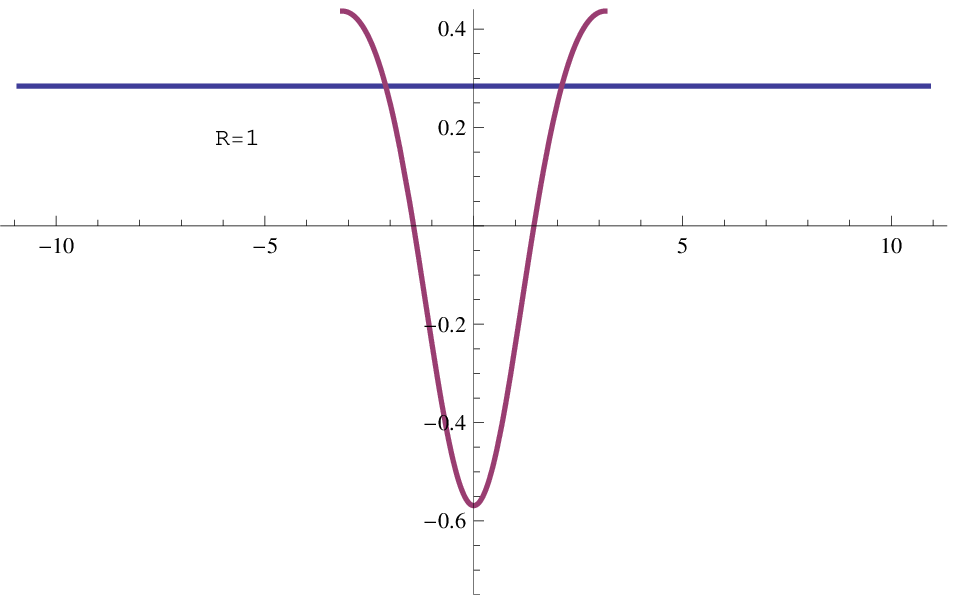}}\\ \  \\
\resizebox{2.3in}{1.4in}{\includegraphics{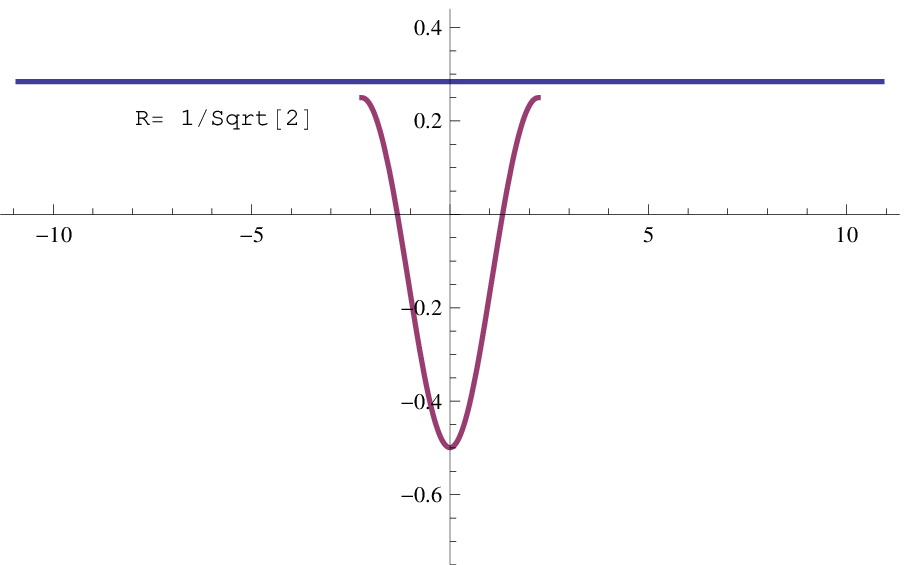}}\ \ \ \ \ \ \ \ \ \
\resizebox{2.3in}{1.4in}{\includegraphics{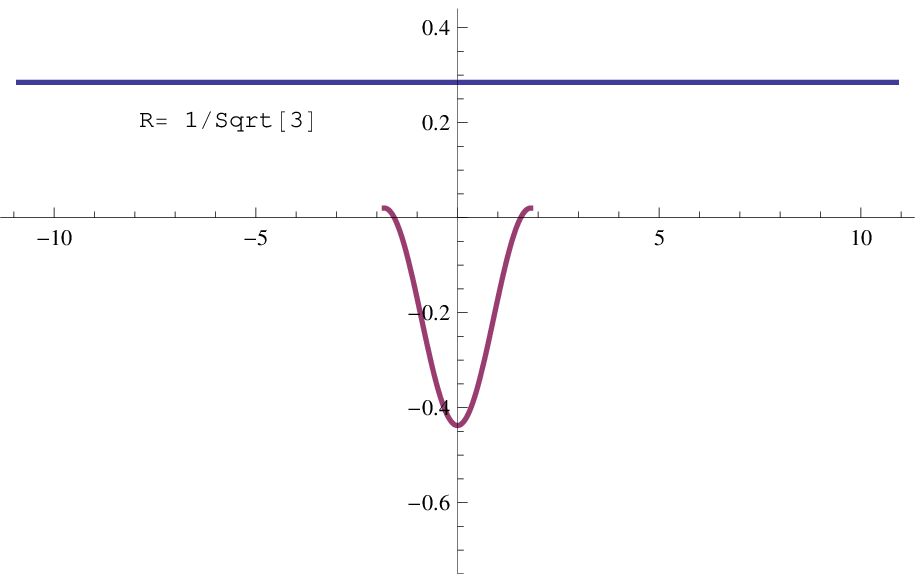}}\\ \ \\
\resizebox{2.3in}{1.4in}{\includegraphics{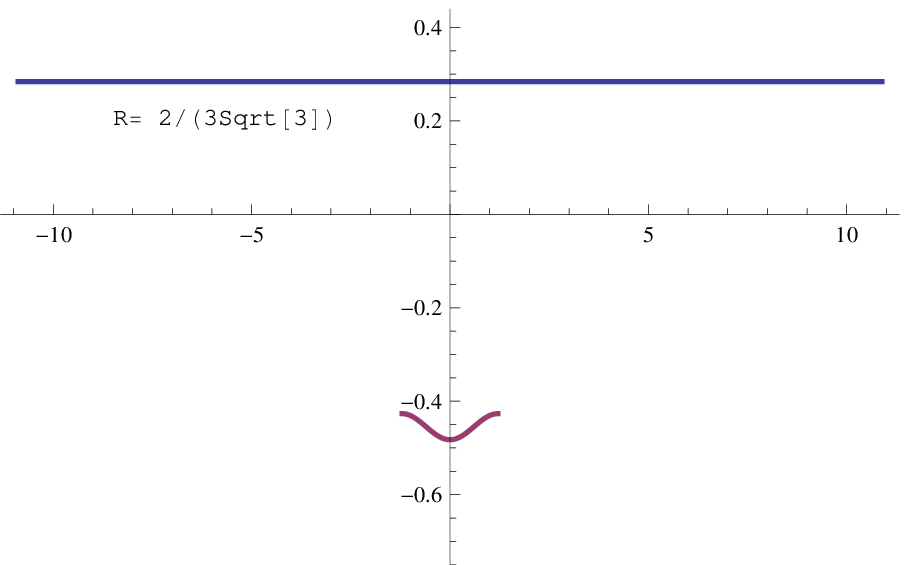}}\ \ \ \ \ \ \ \ \ \ 
\resizebox{2.3in}{1.4in}{\includegraphics{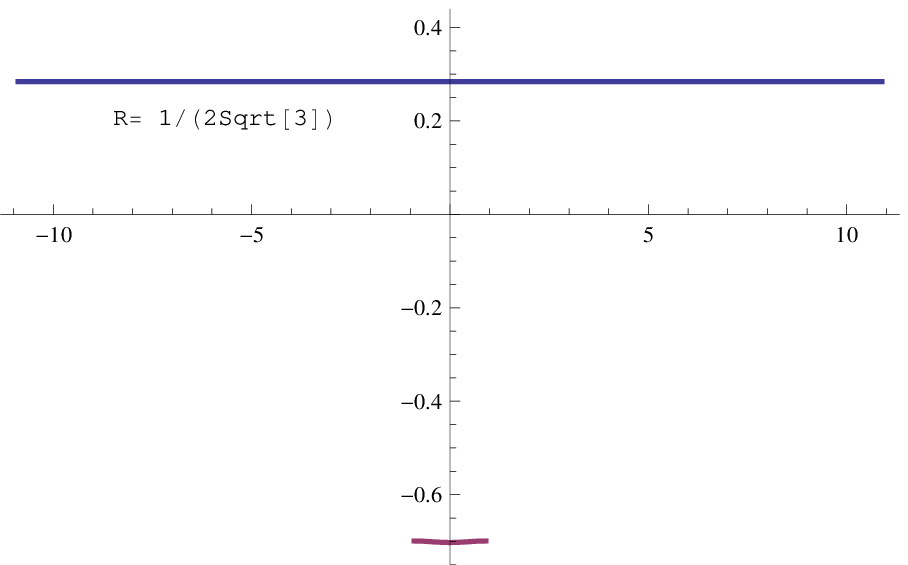}}
\end{center}
\caption{\label{fig:profiles} Lump profiles plotted, starting 
from the top left, for radii $R=2\sqrt{3},
\sqrt{3},\sqrt{2},1,\frac{1}{\sqrt{2}},\frac{1}{\sqrt{3}},
\frac{2}{3\sqrt{3}}$ and $\frac{1}{2\sqrt{3}}$. }
\end{figure}

Now we are ready to plot the tachyon profile, which we show for 
various values of the compactification radius $R$ in figure 
\ref{fig:profiles}. For $R>>1$, the profile takes a fixed form with negligible 
differences for different values of $R$. Similar behavior was observed for 
Siegel gauge lumps in \cite{MSZ}. The largest radius shown in figure 
\ref{fig:profiles} is $R=2\sqrt{3}\approx 3.5$, which 
is already representative of the profile at larger radius. As $R$ approaches
$1$ from above, the tail of the lump begins to feel the finite size of the 
box, but the core is mostly unaffected. As we cross the marginal 
threshold $R=1$ we enter uncharted territory since (at the time of 
writing) no solution has been identified in level truncation. We find that 
for $R<1$ the depth of the lump rapidly decreases, while its spatial average 
falls to negative values. The smallest radius shown in figure 
\ref{fig:profiles} is $R=\frac{1}{2\sqrt{3}}$, 
where the profile is almost completely flat with 
$t(x)\approx -.7$. This behavior can also be seen in the explicit form of 
the tachyon coefficients, as shown in table \ref{tab:coef}. 

Curiously, negative values of the tachyon correspond to falling down the 
unbounded side of the tachyon effective potential, which should contribute 
negatively to the energy. This makes it difficult to understand where the 
positive energy of the lower dimensional D-brane comes from when $R<1$. 
Currently we cannot provide insight into this question, as it may require a 
(possibly high level) analysis of the solution in 
level truncation. The first positive energy solution in string field theory 
was identified in level truncation quite recently \cite{Ising}, in the 
context of a systematic study of classical solutions describing Ising 
model boundary conditions. This result provides evidence that positive
energy solutions can be understood in a controlled manner in level 
truncation.\footnote{The $\s$-brane solution constructed in \cite{Ising} 
has positive tachyon coefficient $\approx .1454$. This result runs counter to 
the expectation derived from \eq{sol}, which always produces a negative 
tachyon coefficient for higher energy solutions. We thank M. Schnabl for 
sharing this piece of numerical data.}

Since we have exact formulas, there are many features of the tachyon profile 
that can in principle be studied analytically. One particularly interesting 
property is the fact that the profile is basically fixed for sufficiently large
radius. From the perspective of the coefficients $t_n$ this is quite 
surprising, since they vary substantially with $R$ well past the point where 
the profile is stable. Since the profile rapidly decays away from the core, 
one way to understand this phenomenon is that the profile at finite (but 
sufficiently large) radius is approximately equal to a periodic sum of the 
profiles at infinite radius. In fact, we claim that this property is 
exact: The profile at finite radius $t(x,R)$ 
can be written in terms of the profile at infinite radius $t(x,\infty)$ with
the formula
\begin{equation}t(x,R)=t_0(\infty)+\sum_{n\in \mathbb{Z}}
\Big(t(x+2\pi R n,\infty)-t_0(\infty)\Big),\label{eq:add}
\end{equation}
where $t_0(\infty)$ is the zero mode coefficient \eq{t0} evaluated at 
$R=\infty$. We give a proof in appendix \ref{app:add}. Note that this is 
a special feature of the solution we have been working with, and does not 
hold in a more general gauge. 

\begin{table}
{\footnotesize
\begin{center}
\begin{tabular}{|c|ccccc|}
\hline
$R$   & $+t_0$   & $-t_1$      & $-t_2$  &   $-t_3$ &    $-t_4$ \\
\hline
$ 3\sqrt3$  & $ 0.229663$  & $0.0566884$  & $0.0604255$  
& $0.0617708$ &$0.0581978$  \\
\hline
$ 2\sqrt3$  & $0.202297$  & $0.0878951$  & $0.0926562$  
& $0.0818861$ & $0.0592841 $    \\
\hline
$ \sqrt3$  & $ 0.120199$  & $ 0.185312$ & $0.118568$  
& $0.0381503$  & $0.00775386$ \\
\hline
$\sqrt2$ & $0.0832971$  & $0.220203$ & $0.0936014$  
& $0.0165937$  & $0.00174037$            \\
\hline
$1.2$  & $0.047399$  & $0.242197$ & $0.0643166$  
& $0.00615465$ & $0.000323713$  \\
\hline
$1.1$ & $0.025854$ & $0.248766$ & $0.0485287$  
& $0.0031664$ &  $0.00010647 $    \\
\hline
\hline
$1$ & $0$ & $0.250000$  & $0.0330126$  & $0.00134221$  
& $2.53129\times10^{-5}$ \\
\hline
\hline
$1/1.1$ &  $-0.0284394$  & $0.243616$  & $0.0205228$  & $0.000485494$ & 
$4.57246\times10^{-6}$ \\
\hline
$1/1.2$ & $-0.0568788$  & $0.230414$  & $0.0121757$  & $0.000162893$ & 
$7.18534\times10^{-7} $      \\
\hline
$1/\sqrt{2}$ & $-0.1178$  & $0.187203$  & $0.00348075$  & $0.0000121516$ &    
$8.48188\times10^{-9} $  \\
\hline
$1/\sqrt3$ & $-0.208191$  & $0.114451$ & $0.000404475$ & 
$1.33399\times10^{-7}$ &    $3.50603\times10^{-12} $   \\
\hline
$1/2\sqrt3$ & $-0.700776$  &  $0.000808949$ & $7.01207\times10^{-12}$ 
& $1.82338\times10^{-24}$ &    $1.12733\times10^{-41} $   \\
\hline
$1/3\sqrt3$  & $-1.19336$  & $4.00196\times10^{-7}$  & 
$2.73507\times10^{-24}$  & $7.5909\times10^{-52}$ &    
$6.925739\times10^{-90} $   \\
\hline
\end{tabular}
\end{center}
}
\caption{\label{tab:coef} List of tachyon coefficients $t_0,...,t_4$ for 
various values of the compactification radius. Note that for $R<1$ the 
coefficients for the nonzero harmonics rapidly become negligibly small, while
the zeroth harmonic becomes increasingly negative. Note also that at $R=1$ we 
obtain $t_1=-\frac{1}{4}$, which means that the coefficient of the marginal 
operator $\cos (X^1/R)$ in the solution is $-1/2$. For solutions describing 
nonsingular marginal deformations, the coefficient of the marginal operator 
in the Fock space expansion is equal to the marginal parameter describing 
the background in BCFT. Thus we consistently find that the solution describes 
the cosine marginal deformation at the critical value 
$\lambda=-\frac{1}{2}$ where the boundary condition is Dirichlet.}
\end{table}

\section{Multiple D-brane Solutions}

In this section we will discuss the construction of backgrounds involving 
more than one D-brane. We will first describe a solution representing multiple
D$(p-1)$-branes in the string field theory of a single D$p$-brane. Then we 
generalize to find a solution describing multiple copies of the perturbative 
vacuum. 

In the last section we found a solution for a single D$(p-1)$-brane on a 
circle of radius $R$. This automatically gives a solution for a pair of 
D$(p-1)$-branes on a circle of radius $2R$, one located at $X^1=0$ and 
the other located at $X^1=2\pi R$. Due to the remarkable property \eq{add}, 
the ``double lump'' profile for this two D-brane system is simply the sum 
of the ``single lump'' profiles centered at $0$ and at $2\pi R$:
\begin{eqnarray}t_{\mathrm{double\ lump}}(x,2R) \lineup = t(x,R),\nonumber\\
\lineup = t(x,2R) + t(x+2\pi R,2R)-t_0(2R).
\end{eqnarray}
One might guess that if the D-branes are at positions $a$ and $b$, we should 
likewise sum the lump profiles centered at $a$ and $b$. This 
immediately suggests that the solution (around the tachyon vacuum) is simply 
the sum of the solutions creating a D$(p-1)$-brane at position $a$ and 
a D$(p-1)$-brane at position $b$:
\begin{equation}\Phi=-\S_a\Psi_\tv\Sb_a - \S_b\Psi_\tv\Sb_b.
\label{eq:ab}\end{equation}
Remarkably, this turns out to be correct. To see this, it is convenient 
to assemble the bcc operators at $a$ and $b$ into row and column vectors 
\begin{equation}{\bm\s} \equiv \big(\s_a\ \ \ \s_b\big);
\ \ \ \ {\bm\sb}\equiv
\left(\begin{matrix}\sb_a\\ \sb_b\end{matrix}\right).\end{equation} 
The row and column have the obvious interpretation  of ``creating'' 
Chan-Paton factors out of a boundary condition where they are absent. 
Building ${\bm\S}$ and ${\bm\Sb}$ from ${\bm\s}$ and ${\bm\sb}$, we can 
write the solution as
\begin{equation}\Phi = -{\bm\S}\Psi_\tv{\bm\Sb}.\label{eq:bmS}\end{equation}
For this to satisfy the equations of motion, ${\bm \sb\bm\s}$ must be equal 
to the $2\times 2$ identity matrix. Computing we find 
\begin{equation}\bm\sb\bm\s = \left(\begin{matrix}1 & \sb_a\s_b\, \\
\,\sb_b\s_a & 1\end{matrix}\right).\end{equation}
The diagonals work correctly because $\sb_a\s_a=\sb_b\s_b=1$. To understand
what happens with the off-diagonal elements, note that the leading term in 
the OPE between $\sb_{\ND,a}$ and $\s_{\ND,b}$ must be proportional to a bcc 
operator which shifts the Dirichlet boundary condition from $a$ to $b$. 
If $a\neq b$ (modulo the circumference of the circle), this operator must have 
positive conformal weight, which means that the leading singularity in the 
$\sb_{\ND,a}$-$\s_{\ND,b}$ OPE must be {\it less severe} than $s^{-1/8}$. 
But the Neumann-Dirichlet twist operators are always accompanied by the 
timelike Wilson line, and the OPE of these bcc operators vanishes 
as $s^{1/8}$. Therefore, $\sb_a\s_b=0$ and
\begin{equation}\bm\sb\bm\s = \left(\begin{matrix}\,1 & 0\, \\
\,0 & 1\,\end{matrix}\right).\end{equation}
Therefore \eq{bmS} is a solution. It is clear that this generalizes to any 
number of non-coincident D$(p-1)$-branes by simply adding more entries into 
the row and column vectors.

However, this misses the important case when the D-branes are coincident. 
We cannot use the same bcc operators to describe all D-branes in this case, 
since then $\bm\sb\bm\s$ will be a matrix of ``ones'' rather than the 
identity matrix. However, there are many choices of $\s,\sb$ which implement
the same change of boundary condition. In the examples discussed so far, 
we have chosen $\s,\sb$ in such a way that the spacelike factor has the 
lowest possible conformal weight. But we can also consider ``excited'' bcc 
operators. For example, we can build the lump solution using
\begin{equation}\s'(s) = \frac{i}{\sqrt{2}}
\d X^2\, \s_\ND e^{i\sqrt{17/16}\,X^0}(s);
\ \ \ \ \sb'(s)=\frac{i}{\sqrt{2}}\d X^2\, \sb_\ND e^{-i\sqrt{17/16}\,X^0}
(s),\label{eq:lumpbcc2}
\end{equation}
where $X^2$ is a free boson orthogonal to $X^1$. The computation of 
observables indicates that the lump solution built from $\s',\sb'$ is 
physically identical to the previous lump solution built 
from $\s,\sb$ in \eq{lumpbcc}.\footnote{In principle, the lump solutions 
built from \eq{lumpbcc2} and \eq{lumpbcc} should be gauge equivalent. Since
they are already in the same gauge, this indicates that the gauge 
condition does not define the solution uniquely. This phenomenon was already
observed in \cite{Schnabl}, where a 1-parameter family of 
solutions for the perturbative vacuum was found in Schnabl gauge. We thank 
M. Schnabl for discussions on this point.} In fact, these two sets of bcc 
operators have vanishing OPE, which means that the row and column vectors
\begin{equation}{\bm\s}= \big(\s\ \ \ \s'\big);
\ \ \ \ \ \ 
{\bm\sb}=\left(\begin{matrix}\sb\\ \sb'\end{matrix}\right),
\end{equation}
define a solution for a coincident pair of D$(p-1)$-branes.

By the same mechanism, we can construct a ``double brane'' solution 
describing two copies of the perturbative vacuum. Here there is no change of 
boundary condition, so we simply construct $\bm\s,\bm\sb$ using the 
primaries of $\BCFT_0$. For example, if $\BCFT_0$ is made from free bosons,
we can define two sets of ``bcc operators'':
\begin{eqnarray}
\s_1(s) \lineup = \frac{i}{\sqrt{2}}\d X^1 e^{iX^0}(s);\ \ \ \ 
\sb_1(s) = \frac{i}{\sqrt{2}}\d X^1 e^{-iX^0}(s);\nonumber\\
\s_2(s) \lineup = \frac{i}{\sqrt{2}}\d X^2 e^{iX^0}(s);\ \ \ \ 
\sb_2(s) = \frac{i}{\sqrt{2}}\d X^2 e^{-iX^0}(s).\label{eq:double}
\end{eqnarray}
Defining row and column vectors 
\begin{equation}{\bm\s}= \big(\s_1\ \ \ \s_2\big);
\ \ \ \ \ \ 
{\bm\sb}=\left(\begin{matrix}\sb_1\\ \sb_2\end{matrix}\right),
\end{equation}
we have $\bm\sb\bm\s=\mathbb{I}_{2\times 2}$, and the solution creates 
two copies of the perturbative vacuum. Actually, this is probably the 
simplest nontrivial solution discussed so far. The $n$-point functions of 
$\bm\s$ and $\bm\sb$ can be computed by elementary means, and even the 
Schnabl-gauge solution \eq{Sch2} can plausibly be studied in a fairly 
explicit manner. Note that, contrary to some expectations, this multibrane 
solution is not formulated within the universal sector. A different 
approach to multibrane solutions, advanced in \cite{MS} and further explored 
in \cite{Dtak,MNT,Masuda,HataKojita1,HataKojita2,HataKojita3,Arroyo1,Arroyo2}, 
requires only universal states generated by $K,B$ and $c$. However, 
the solution is quite singular and an adequate regularization has not been 
found. Also, it is unclear in this approach how non-abelian gauge bosons 
emerge in the spectrum of excitations.

Having discussed a few explicit examples, let us outline the general 
construction. Suppose that, starting from $\BCFT_0$, we want to describe 
a system of $N$ D-branes described by boundary conformal field theories 
$\BCFT_i$ for $i=1,...,N$. We need $N$ bcc operators 
\begin{equation}\s_i(s) = \s_{*,i}\,e^{i\sqrt{h_i}X^0}(s);\ \ \ \ 
\sb_i(s) = \sb_{*,i}\,e^{-i\sqrt{h_i}X^0}(s),
\end{equation}
where $\s_{*,i},\sb_{*,i}$ are primaries of weight $h_i$ which act as the 
identity operator in the time direction and change the 
boundary condition from $\BCFT_0$ to $\BCFT_i$ in the spatial ($c=25$) 
directions. If the bcc operators satisfy
\begin{equation}\lim_{s\to 0}\sb_i(s)\s_j(0) = \delta_{ij},\ \ \ \ (s>0),
\label{eq:multis}\end{equation}
then row and column vectors
\begin{equation}
\bm\s = \big(\s_1\ \ \ ...\ \ \  \s_N\big);\ \ \ \ 
\bm\sb = \left(\begin{matrix}\sb_1\\ \vdots \\ \sb_N\end{matrix}\right),
\label{eq:genrowcol}
\end{equation}
define a solution for the desired multiple D-brane system. The orthogonality
condition \eq{multis} is nontrivial. In general the $\sb_{*,i}$-$\s_{*,j}$
OPE takes the form 
\begin{equation}
\sb_{*,i}(s)\s_{*,j}(0) = \frac{1}{s^{h_i+h_j-h_{ij}}}\s_{*,ij}(0)+
\mathrm{less\ singular},\ \ \ \ (s>0),
\end{equation}
where the leading term is proportional to a boundary condition changing 
operator $\s_{*,ij}$ relating $\BCFT_i$ with $\BCFT_j$ with dimension $h_{ij}$.
The orthogonality condition is satisfied provided the conformal weights 
of the operators in this OPE satisfy the bound
\begin{equation}|\sqrt{h_i}-\sqrt{h_j}|<\sqrt{h_{ij}},\ \ \ \ 
(i\neq j).\end{equation}
We are not certain whether this inequality poses a limitation on the possible 
multiple D-brane systems that can be constructed by our method. In 
the case where the $c=25$ theory is described by free bosons, we have 
confirmed that it is possible to create an arbitrarily large number of 
copies of the perturbative vacuum by choosing $\s_i$s consistent with 
this bound.\footnote{One of many possible choices is 
$\s_{j,*}=\sb_{j,*}=\frac{i}{\sqrt{2}}\d X^2 p_{j^2}^{(1)}$ where 
$p_{j^2}^{(1)}$ are the (properly normalized) zero momentum primaries in 
the $X^1$ BCFT of weight $j^2$.}

In noncommutative field theories \cite{GMS} and vacuum string field 
theory \cite{bef,VSFT,VSFTsol1,SSF}, there is a close relation
between multiple D-brane systems and higher rank projectors. While the solution
\eq{sol} is not a star algebra projector, there is a natural way to associate
a star algebra projector to any classical solution in open string field 
theory \cite{Ellwood,Integra}. The construction goes as follows. Consider a 
``singular'' gauge transformation defined by the string 
field\footnote{We can in principle compute a 
projector given $U=Q_\Psi b$ for any ghost number $-1$ state $b$, but 
the choice \eq{U} simplifies the calculation.}
\begin{eqnarray}U \lineup  = Q_\Psi\left(\frac{B}{1+K}\right),\nonumber\\
\lineup = 1-\bm\S \frac{1}{1+K}\bm\Sb.\label{eq:U}\end{eqnarray}
Formally, $U$ is a gauge parameter defining a (reducible) gauge 
transformation from the solution to itself: 
\begin{equation}\Psi = U^{-1}(Q+\Psi)U,\ \ \ \ \ (\mathrm{formally}).
\end{equation}
But in reality this gauge transformation is singular. To see why, note 
that the definition of $U$ together with the (presumed)
existence of $U^{-1}$ implies that the identity string field is 
trivial in the $Q_\Psi$ cohomology:
\begin{equation}1 = Q_\Psi\left(U^{-1}\frac{B}{1+K}\right),\end{equation} 
which would mean that the solution supports no open string excitations. 
Therefore, if the solution is not the tachyon vacuum, we are forced to 
conclude that $U$ is not invertible. If we think of $U$ as an operator 
on the space of half string functionals \cite{VSFTsol1,SSF}, it should 
have a kernel. The projector onto the kernel is called the 
{\it characteristic projector}, and if $U=1-X$ we can compute the 
characteristic projector from the limit
\begin{equation}X^\infty = \lim_{N\to\infty}X^N.\end{equation}
Plugging in \eq{U} we find
\begin{eqnarray}
X^\infty \lineup = \bm\S\,\Omega^\infty\,\bm\Sb,\nonumber\\
\lineup = \S_1\,\Omega^\infty\,\Sb_1 + \,.\,.\,.\,+\S_N\,\Omega^\infty\,\Sb_N,
\end{eqnarray}
where $\Omega^\infty$ is the sliver state \cite{RZ_wedge,proj}, and 
in the second 
step we expanded $\bm\S,\bm\Sb$ out into components $\S_i,\Sb_i$ creating 
the boundary condition of each constituent D-brane. The sliver state 
factorizes into a wavefunctional on the left and right halves of the open 
string, and therefore can be interpreted as a rank one projector 
\cite{VSFTsol1,SSF}. Therefore, $X^\infty$ is a sum of 
rank one projectors carrying the boundary condition of each constituent 
D-brane. Moreover, since 
\begin{equation}\Sb_i\S_j = \delta_{ij},\end{equation}
the projectors are $*$-orthogonal. Therefore, for a system of $N$ D-branes, 
the characteristic projector (formally) has rank $N$.  The picture that 
emerges strongly resembles the boundary conformal field theory construction 
of D-branes in vacuum string field theory \cite{bef}, where the equations 
of motion are solved by adding sliver states with appropriately deformed 
boundary conditions. There are interesting differences, however. 
In \cite{bef} the projectors are rendered $*$-orthogonal by the nontrivial 
conformal weight of the boundary condition changing insertions, which under 
star multiplication produce a vanishing factor due to a singular conformal 
transformation. In our construction, the matter insertions carry vanishing 
conformal weight, and the singular multiplication of sliver states is not 
essential. Rather, the projectors are $*$-orthogonal because the boundary 
insertions themselves are already $*$-orthogonal.

\section{Conclusion}

To summarize, the solution takes the form
\begin{equation} \Psi_\tv-\S\Psi_\tv\Sb, \label{eq:consol}\end{equation}
where $\Psi_\tv$ is the tachyon vacuum \eq{simple} and $\S$ and $\Sb$ string 
fields which change the open string boundary condition between the 
perturbative vacuum and the D-brane system we wish to describe. The form of 
the solution is easy to grasp. To find a new background, we first condense to 
the tachyon vacuum (the first term), then we ``reverse'' the 
process of tachyon condensation to create the new D-brane system (the second
term). Note, in particular, that $\S\Psi_\tv\Sb$ is the tachyon 
vacuum of the new D-brane system reexpressed (via $\S$ and $\Sb$) in the 
variables of the reference boundary conformal field theory. The solution 
reproduces the physics of the new background in the sense that:
\begin{itemize}
\item The action evaluated on the solution describes the 
difference in tension between the perturbative vacuum and the target 
D-brane system.
\item The solution implies the correct coupling between the new background
and closed string states. 
\item The action expanded around the solution, after a trivial field 
redefinition, is identical to the string field theory formulated 
in the new background. In this sense, the background independence of open 
string field theory is manifest. 
\end{itemize}
These results depend very little on the detailed form of the solution.  
They follow quite generally from the relations 
\begin{equation}\Sb\S = 1;\ \ \ \ \ \ 
Q_{\Psi_\tv}\S=Q_{\Psi_\tv}\Sb = 0,\end{equation}
together with the fact that $\Psi_\tv$ is a solution for the tachyon 
vacuum. This suggests that there may be other solutions which 
share the same basic structure and transcend some limitations in our 
approach.\footnote{Note, for example, that the recent 
solution of \cite{Macc} can be recast in the form \eq{consol} due to its 
formal similarity with the KOS solution. However, the details of \cite{Macc} 
are quite different from the solution discussed here.} Our 
implementation assumes the existence of boundary condition changing
operators of vanishing conformal weight, which for time-independent 
backgrounds we construct by tensoring the background shift with a timelike 
Wilson line of specific magnitude. This construction does not work for 
time-dependent backgrounds. Moreover, this construction excites primaries 
in $\BCFT_0$ which are irrelevant to describing the physics of the new 
background. This can hide symmetries---such as Lorentz invariance---which 
we might prefer to be manifest.

One important question we have not addressed is the behavior of the solution 
in level truncation. This question poses a technical challenge, both because 
the solution is somewhat exotic from the perspective of level truncation---due 
to the excitation of $X^0$ primaries and the existence of higher energy 
configurations---and because the gauge condition 
$\mathcal{B}_{\frac{1}{\sqrt{1+K}},\frac{1}{\sqrt{1+K}}}=0$ produces solutions
which are close to being singular from the perspective of the identity string
field \cite{IdSing}. Indeed, even the tachyon vacuum \eq{simple} gives 
a divergent series for the energy in level truncation, though the series 
can be resummed to give the expected result within less than a percent 
\cite{simple}. The Schnabl gauge solution \eq{Sch2} in theory should be 
a safer starting point for level truncation studies, but the evaluation 
of $2n$-point functions of bcc operators presents a substantial technical 
obstacle. 

In this paper we have focused on the bosonic string, but clearly it would be 
interesting to generalize these results to the superstring. Given the 
central role of the tachyon vacuum \eq{simple} in our construction, we 
expect that the tachyon vacuum of Berkovits superstring field theory, 
recently found in \cite{BerkVac}, will likewise play a central role 
for the superstring. We hope to return to this question soon.

The solution we have found appears to solve several longstanding and 
fundamental problems in string field theory, and, with remarkable simplicity,
demonstrates the power of string field theory to provide a unified 
description of the multitude of backgrounds of first quantized string theory. 
We hope to see exciting developments in the near future.

\bigskip

\noindent {\bf Acknowledgments}

\bigskip

\noindent We would like to thank I. Pesando for helpful discussions 
about Wilson lines and twist fields, and M. Schnabl for discussions 
and comments on a draft of this paper. C.M. thanks L. Bonora and D. Gaiotto 
for interesting conversations and M. Schnabl for hospitality in Prague 
where much of this work was completed. The work of T.E. was supported 
in parts by the DFG Transregional Collaborative Research Centre TRR 33, 
the DFG cluster of excellence Origin and Structure of the Universe. 
The research of CM is funded by a  
{\it Rita Levi Montalcini} grant from the Italian MIUR, and as 
also been supported by the Grant Agency of the Czech Republic, under the 
grant P201/12/G028.

\begin{appendix}

\section{Four point function of twist fields}
\label{app:twist}

In this appendix we discuss the 4-point function of 
Neumann-Dirichlet twist operators for a free boson $X^1$ compactified on a 
circle of radius $R$. Most computations in this paper do not 
require this correlator, but it implicitly appears (for example) in the 
computation of the quadratic term in the equations of motion and the kinetic 
term in the action. (The cubic term implicitly requires the six point 
function). Our main interest in the 4-point function is as a cross-check 
on the OPE \eq{NDOPE}, and as an illustration of the algebraic structure of 
the solution in the context of a correlator which is not completely fixed by 
conformal invariance. 

\begin{figure}
\begin{center}
\resizebox{2.3in}{1.4in}{\includegraphics{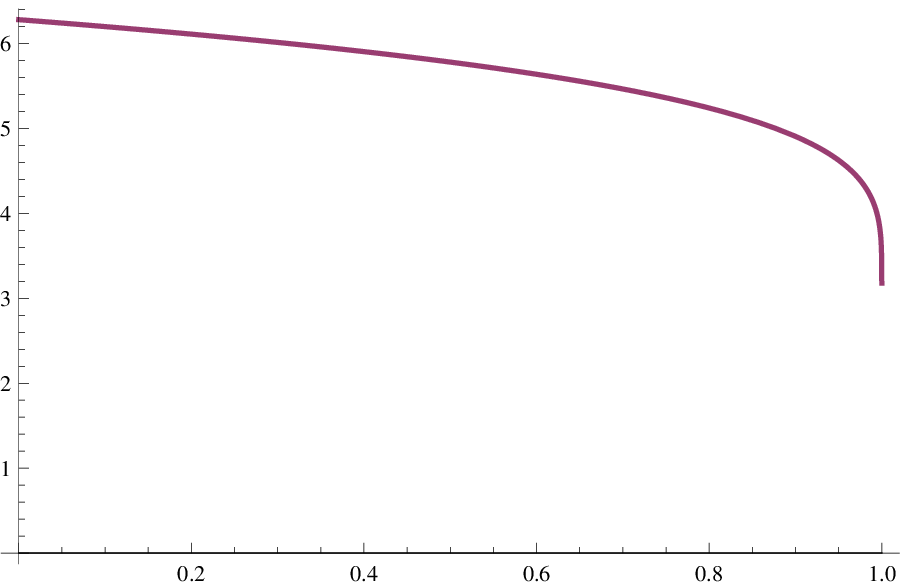}}\ \ \ \ \ \ \ \ \ \ 
\resizebox{2.3in}{1.4in}{\includegraphics{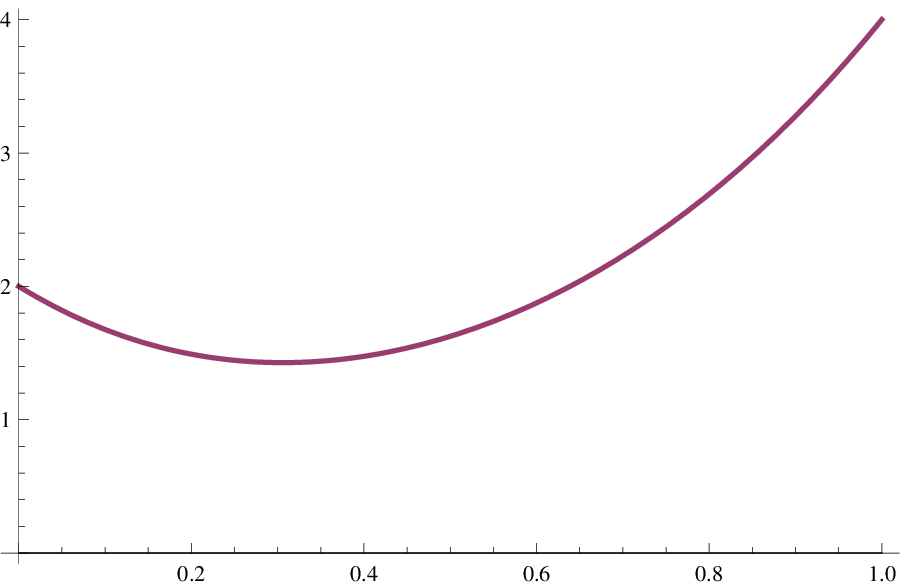}}\\ \  \\
\end{center}
\caption{\label{fig:4pt} The first plot shows the correlator \eq{4ptW} as a 
function of $s\in[0,1]$ when $R=2$. Note that the correlator is 
non-differentiable at $s=1$, and the value is half that at $s=0$. For 
illustrative purposes, in the second plot we show the 4-point 
function of bcc operators for the 2-brane solution \eq{double}. Here 
the value at $x=1$ is twice that at $x=0$, which represents the doubling 
of energy.}
\end{figure}

The complete 4-point function of Neumann-Dirichlet twist fields, including 
instanton corrections from the compactification, was computed in 
\cite{Narain} and takes the form
\begin{equation}
\big\langle I\circ\sb_\ND(0)\s_\ND(1)\sb_\ND(s)
\s_\ND(0)\big\rangle^{X^1}_\mathrm{UHP} = 
\frac{2\pi}{|s(1-s)|^{1/8}}G(s,R),
\label{eq:4pt}\end{equation}
where\footnote{Our notation for elliptic functions follows Gradshteyn 
and Ryzhik \cite{GR}.}
\begin{equation}G(s,R)=\frac{1}{\sqrt{\frac{2}{\pi}K\big(\sqrt{s}\,\big)}}
\,\vartheta_3\left(0,q\big(\sqrt{s}\,\big)^{R^2}\right).\end{equation}
Here $\vartheta_3(0,q)$ is the Jacobi theta function, $K(k)$ the complete 
elliptic integral of the first kind, and $q(k)$ the nome
\begin{equation}q(k) = e^{-\pi\frac{K(\sqrt{1-k^2})}{K(k)}}.
\end{equation}
Tensoring $\s_\ND,\sb_\ND$ with the timelike Wilson line removes the 
singular factor $|s(1-s)|^{-1/8}$:
\begin{equation}
\big\langle \sb(\infty)\s(1)\sb(s)
\s(0)\big\rangle^{\mathrm{matter}}_\mathrm{UHP}
=2\pi\, G(s,R).\label{eq:4ptW}
\end{equation}
We plot this for $s\in[0,1]$ in figure \ref{fig:4pt}. Note that the modular 
property of the theta function,
\begin{equation}\vartheta_3(0,e^{i\pi(-1/\tau)})=(-i\tau)^{1/2}\vartheta_3(
0,e^{i\pi\tau}),\end{equation}
implies that the 4-point function satisfies 
\begin{equation}G(1-s,R)=\frac{1}{R}G\left(s,\frac{1}{R}\right).
\label{eq:mod}\end{equation}
This is obviously related to T-duality. The correlator at radius $R$ and $1/R$
are related by a switch of Neumann and Dirichlet boundary conditions, which 
in effect interchanges $\s_\ND$ and $\sb_\ND$. At the self dual radius, 
$G(s,R)$ is constant:
\begin{equation}G(s,1)=1.\end{equation} 
Note that the points $s=0$ and $s=1$ represent a collision between
$\s$ and $\sb$, where the 4-point function reduces to a 2-point 
function. At these points we find
\begin{equation}G(0,R)=1;\ \ \ \ G(1,R)=\frac{1}{R}.\end{equation}
This is confirms the OPE
\begin{equation}\lim_{s\to0}\sb(s)\s(0) =1;\ \ \ \ \lim_{s\to 0}\s(s)\sb(0)
=\frac{1}{R}.\end{equation}
Thus the ``associativity anomaly'' can be seen explicitly 
in the 4-point function. 

Let us take a closer look at the behavior of the 4-point function 
in the limit $s\to 1$, where we should be able to extract the 
$\s_\ND$-$\sb_\ND$ OPE computed in \eq{NDOPE}. To make the expansion 
somewhat easier, it is convenient to use T-duality \eq{mod} to map from 
$s=1$ to $s=0$. Then using the power series
\begin{eqnarray}\vartheta_3(0,q)\lineup = 1+2q+\mathcal{O}(q^4);\\
\frac{2}{\pi}K(k)\lineup = 1+\frac{k^2}{4}+\frac{9k^4}{64}+\mathcal{O}(k^6);\\
q(k)\lineup = \frac{k^2}{16}+\frac{k^4}{32}+\mathcal{O}(k^6),
\end{eqnarray}
we obtain
\begin{eqnarray}
\lineup \!\!\!\!\!\!\!\!\!\!
\big\langle I\circ\sb_\ND(0)\s_\ND(1)\sb_\ND(1-s)
\s_\ND(0)\big\rangle^{X^1}_\mathrm{UHP} =\nonumber\\
\lineup\ \ \ \ \ \ \ \ \ \ \ \ \ \ \ \ \ \ \ \ \ \ \ \ \ \ \ \ \ 
\frac{1}{s^{1/8}}\frac{2\pi}{R}
+\frac{1}{s^{1/8-1/R^2}}\frac{2\pi\cdot2^{-4/R^2+1}}{R}+\,.\,.\,.\,,\ \ \ \ 
(R>1/\sqrt{2}).\nonumber\\ \label{eq:s1exp}
\end{eqnarray}
The first and second terms represent the contribution from the identity 
operator and the first cosine harmonic, respectively, in the OPE between 
$\s_\ND$ and $\sb_\ND$. The restriction $R>1/\sqrt{2}$ is assumed 
otherwise the second term is subleading to terms of the order $s^{15/8}$, 
which arise from the first Virasoro descendent of the identity. Alternatively,
we should be able to compute \eq{s1exp} by substituting the OPE \eq{NDOPE} 
directly into the correlator:
\begin{eqnarray}
\lineup \!\!\!\!\!\!\!\!\!\!\!
\big\langle I\circ\sb_\ND(0)
\s_\ND(1)\sb_\ND(1-s)\s_\ND(0)\big\rangle_\mathrm{UHP}^{X^1}\nonumber\\
\lineup\ \ \ \ 
 = \frac{1}{s^{1/8}}\frac{1}{R}\big\langle I\circ\sb_\ND(0)\s_\ND(0)
\big\rangle_\mathrm{UHP}^{X^1}
\nonumber\\
\lineup\ \ \ \ \ \ \ \ +\frac{1}{s^{1/8-1/R^2}}\frac{2^{-2/R^2+1}}{R}
\left\langle I\circ\sb_\ND(0)\cos\left(\frac{X^1-a}{R}\right)(1)\s_\ND(0)
\right\rangle_\mathrm{UHP}^{X^1}
+\,.\,.\,.\,,\nonumber\\
\lineup\ \ \ \ \ \ \ \ \ \ \ \ \ \ \ \ \ \ \ \ \ \ \ \ \ \ \ \ \ \ 
\ \ \ \ \ \ \ \ \ \ \ \ \ \ \ \ \ \ \ \ \ \ \ \ \ \ \ \ \ \ \ 
\ \ \ \ \ \ \ \ \ \ \ \ \ \ (R>1/\sqrt{2}).
\label{eq:s1OPEexp}
\end{eqnarray}
Using the 3-point function \eq{3pt} 
\begin{equation}\big\langle I\circ\sb_\ND(0)e^{inX^1/R}(1)\s(0)
\big\rangle_\mathrm{UHP}^{X^1} = 2\pi 2^{-2n^2/R^2} e^{in a/R},
\end{equation}
we find agreement between \eq{s1exp} and \eq{s1OPEexp}.

\section{Additivity of the Lump Profile}
\label{app:add}

In this appendix, we prove that the lump profile for the solution
\eq{sol} compactified on a circle of radius $R$ is a periodic sum of the 
uncompactified lump profile. The profile at radius $R$ is given by 
\begin{equation}t(x,R) = t_0(R) + \sum_{n\in\mathbb{Z}-\{0\}}\frac{1}{R}
2^{-2n^2/R^2}g(n^2/R^2)e^{in x/R},\label{eq:profR}\end{equation}
where $t_0(R)$ is given by \eq{t0}. In the 
limit $R\to\infty$, the sum turns into an integral
\begin{equation}t(x,\infty)=t_0(\infty)+\int_{-\infty}^\infty dk\, 
2^{-2k^2}g(k^2)e^{ikx}.\label{eq:Tinf}
\end{equation}
Our goal is to establish
\begin{equation}
t(x,R) = t_0(\infty) 
+\sum_{n\in \mathbb{Z}}\Big(t(x+2\pi R n,\infty)-t_0(\infty)\Big).
\end{equation}
Substituting \eq{Tinf}, we should have
\begin{equation}
t(x,R) = t_0(\infty)+\sum_{n\in \mathbb{Z}}\int_{-\infty}^\infty dk\, 
2^{-2k^2}g(k^2)e^{ik(x+2\pi R n)}.
\end{equation}
Performing the sum over Fourier harmonics gives a ``Dirac comb'' of delta 
functions:
\begin{equation}\sum_{n\in \mathbb{Z}}e^{ik(x+2\pi R n)}=
\sum_{n\in\mathbb{Z}}\frac{1}{R}\delta(k-n/R).\end{equation}
Evaluating the integral then gives
\begin{equation}
t(x,R) = t_0(\infty) + \frac{1}{R} \lim_{h\to 0}g(h)
+\sum_{n\in\mathbb{Z}-\{0\}}\frac{1}{R}
2^{-2n^2/R^2}g(n^2/R^2)e^{in x/R}.
\end{equation}
This is almost the expected lump profile. All we have to do is show that the 
zero mode works correctly. This requires  
\begin{equation}\lim_{h\to 0}g(h)=-t_0(\infty)\approx -.2844.
\label{eq:exp}\end{equation}
From the form of $g(h)$ given in equations \eq{g1} and \eq{g2}, it is clear 
that the $h\to 0$ limit is determined by the formula
\begin{equation}\lim_{h\to 0} h f(s)^{h-1} =\frac{1}{f'(0)}\delta(s),
\label{eq:qlim}\end{equation}
where $f(s)$ is a function that vanishes at $s=0$ and $f'$ is the first 
derivative. Plugging into \eq{g1} and \eq{g2}, the integration 
over $s$ disappears against the delta function, and the remaining expression 
turns out to be (minus) the tachyon coefficient of the tachyon vacuum, 
as required by \eq{exp}. There is a schematic way to understand 
why this works. The integration variable $s$ in equations \eq{g1} and \eq{g2}
represents the Schwinger parameter for the factor 
\begin{equation}\d\s\frac{1}{1+K}\sb,\end{equation}
which appears in the KOS solution. The formula \eq{qlim} effectively 
says that this factor is replaced by the identity string 
field in the $h\to 0$ limit. Thus the KOS solution becomes (minus) 
the tachyon vacuum: 
\begin{equation}- c\d\s \frac{B}{1+K}\sb (1+K)c\frac{1}{1+K}
\ \longrightarrow\ 
-c(1+K)Bc\frac{1}{1+K},
\end{equation}
and the tachyon coefficient is correspondingly that of the tachyon 
vacuum.

\end{appendix}

\end{document}